\documentclass[useAMS,usenatbib]{mn2e}

\usepackage[T1]{fontenc}
\usepackage{rotfloat}
\usepackage{amssymb, amsmath}
\usepackage{units}
\usepackage{graphicx,epstopdf}

\newcommand{\kms}{km~s$^{-1}$}
\newcommand{\mch}{M_{\mathrm{Ch}}}

\title[3-D morphology of Ia SNRs]
  {Three-dimensional numerical investigations of the morphology of type Ia SNRs}
\author[D. C. Warren and J. M. Blondin]
  {Donald~C.~Warren$^1$, John~M.~Blondin$^1$ \\
  $^1$Physics Department, North Carolina State University, Raleigh, NC 27695-8202, USA}
\date{Released 2012 Xxxxx XX}

\pagerange{\pageref{firstpage}--\pageref{lastpage}} \pubyear{2012}

\def\LaTeX{L\kern-.36em\raise.3ex\hbox{a}\kern-.15em
    T\kern-.1667em\lower.7ex\hbox{E}\kern-.125emX}

\begin{document}

\label{firstpage}

\maketitle

\begin{abstract}
We explore the morphology of Type Ia supernova remnants (SNRs) using three-dimensional
hydrodynamics modeling and an exponential density profile. Our model distinguishes ejecta from the
interstellar medium (ISM), and tracks the ionization age of shocked ejecta, both of which allow for
additional analysis of the simulated remnants. We also include the adiabatic index $\gamma$ as a
free parameter, which affects the compressibility of the fluid and emulates the efficiency of cosmic
ray acceleration by shock fronts. In addition to generating 3-D images of the simulations, we
compute line-of-sight projections through the remnants for comparison against observations of
Tycho's SNR and SN 1006. We find that several features observed in these two remnants, such as the
separation between the fluid discontinuities and the presence of ejecta knots ahead of the forward
shock, can be generated by smooth ejecta without any initial clumpiness. Our results are consistent
with SN 1006 being dynamically younger than Tycho's SNR, and more efficiently accelerating cosmic
rays at its forward shock. We conclude that clumpiness is not a necessary condition to reproduce
many observed features of Type Ia supernova remnants, particularly the radial profiles and the
fleecy emission from ejecta at the central region of both remnants.
\end{abstract}


\section{INTRODUCTION}

SN 1572, visible to the unaided eye on Earth, is now linked to the name of its most accurate
observer, Tycho Brahe. Its remnant was rediscovered in radio frequencies almost four centuries later
\citep{HH52,BE57}, and has since been observed throughout the electromagnetic spectrum. SN 1006 was
probably significantly brighter than SN 1572 to observers \citep{KT08,Wink03} around the globe when
it first appeared, but its remnant was not identified until \citet{GM65}. Both supernova remnants
(hereafter SNRs) are resolvable to several hundred pixels across their diameters with the
\emph{Chandra} X-ray Observatory, and confirmation that the original supernovae were of type Ia
\citep{Baade45,KT08,Wu83} make them ideal laboratories to test theories of supernova and remnant
evolution.

Detailed observations have shown both Tycho's SNR and SN 1006 to be rich in structure. Of the
numerous features observed in Tycho \citep{Sew83,DvBS91,Reyn97,Warr05} and SN 1006 (see \citet{CC08}
and references therein), three aspects of the remnants' morphology are relevant to this paper. One
is the nature of the clumpy, ``fleecy'' thermal emission located throughout Tycho and SN 1006, long
since identified as ejecta in Tycho's case \citep{HSS86} and that could be caused by the
Rayleigh-Taylor instability. Another is the limb brightening effect, which can be used to identify
the reverse shock but which is less pronounced than expected, or even absent, in many places around
Tycho and SN 1006. The third is the radial structure of the remnants; that is, identifying the
radial locations of the forward shock, contact discontinuity, and reverse shock and their projected
positions.

It has long been argued that SNRs would be subject to fluid instabilities such as the 
Rayleigh-Taylor \citep{Gull73} and the Kelvin-Helmholtz shear instability \citep{Fryx91,CBE92}.
Swept-up interstellar material (ISM) decelerates the denser ejecta behind the contact discontinuity,
and fluctuations within the ejecta or ISM seed the Rayleigh-Taylor instability. Early treatment of
this problem was either analytical or limited to one-dimensional numerical simulations looking for
Rayleigh-Taylor-unstable zones in a radial profile \citep{Gull73, DEJR89}. Nonlinearity and
asymmetry limit the effectiveness of either approach, but advances in computing have since allowed
SNRs and relevant instabilities to be modeled in multiple dimensions \citep{Hach90,Fryx91,CBE92}.
These results were expanded upon by other authors for a variety of other scenarios, such as ejecta
clumping \citep{Orl12}, overdense ejecta ``bullets'' \citep{WC01}, underdense ejecta ``bubbles''
\citep{BBR01}, efficient particle acceleration at the shock fronts \citep{BE01,FD10}, SNR
interactions with a moving ISM \citep{VV06}, or asymmetries in the progenitor star and its
environment \citep{BLC96,Vigh11}.

SNRs expand and interact in three dimensions, but are observed in only two. Thus any comparison
against observations must include some projection of the data, leading to projection effects such as
limb brightening or overemphasis of irregular surfaces. \citet{FD10} present a three-dimensional
simulation of one octant of a remnant, including line-of-sight projections of radiating material.
Images both with and without emission from shocked ISM can be found in that paper, and show
developed ejecta structures similar to those in Tycho. They do not, however, accurately capture the
weak limb brightening effect observed by \citet{Warr05}; additionally, they find the reverse shock
closer to the forward shock than observations suggest is the case for Tycho. The work of
\citet{Orl12}, performed over $4\pi$ steradians, uses volume renderings of the evolved remnants to
match the radial structure of SN 1006. Inclusion of ejecta clumps alters the corrugation of the
reverse shock, with the amount of disruption increasing as the size of the clumps increases. It also
allows Rayleigh-Taylor structures generated by these clumps to reach and perturb the forward shock.

The primary question addressed by this paper is whether fluid instabilities alone are sufficient to
explain the morphology of Tycho's SNR and SN 1006 as observed by \emph{Chandra}. All three features
of interest in these SNRs -- the proximity of the contact discontinuity to the forward shock in
projection, the fleecy nature of the central emission, and the indistinct limb brightening in some
locations -- would be affected by fluid instabilities at the contact discontinuity. Projection of a
three-dimensional shell structure to a two-dimensional surface overemphasizes deviations from
spherical symmetry, such as Rayleigh-Taylor fingers; this leads to errors in estimating the
locations of the contact discontinuity and reverse shock. In particular, it has recently been
proposed \citep{Orl12} that clumpy ejecta is necessary to explain these features. We present in this
paper complete three-dimensional simulations of a type Ia supernova remnant, covering the full
angular range to allow unrestricted growth of Rayleigh-Taylor instabilities in all directions,
without preferential treatment of certain wavenumbers.

In \S\ref{sec:The-exponential-model} of this article, we explain the model used in the simulations.
In \S\ref{sec:Numerical-models} we describe the initial conditions of the simulations and present
results from the runs. In \S\ref{sec:Observational-implications} the three-dimensional simulations
are projected into a plane and compared against observations of Tycho's SNR and SN 1006. We discuss
features of the projections in \S\ref{sec:Discussion}, in particular our estimation of the dynamical
age of the two remnants. We conclude in \S\ref{sec:Conclusions}.

\section{THE EJECTA PROFILE}\label{sec:The-exponential-model}

\citet{DC98} compared several models for the ejecta profile of a type Ia supernova -- constant, power
law, and a new exponential profile for ejecta density -- against several hydrodynamical models for
type Ia explosion mechanisms and early spectral evolution. They found that the exponential density
profile fit the models under consideration to a higher degree of accuracy than did the constant or
power law profiles. In particular, the authors noted a good fit with the successful W7 model of
\citet{NTY84} -- see Figure 1 of \citet{DC98} -- and found that the profile began steep but
flattened out over time. We adopt the exponential model developed in that paper, which uses a
spherically symmetric density profile
\begin{equation}
\rho_{\mathrm{SN}}=At^{-3}e^{-v/v_{e}}.\label{eq1}\end{equation}
Assuming that $r=vt$ at $t=t_{0}$, the constants $A$ and $v_{e}$ are found to be
\begin{equation}
A=\frac{6^{3/2}}{8\pi}\frac{M_{e}^{5/2}}{E^{3/2}}=7.67\times10^{6}\,\left(\frac{M_{e}}{\mch}\right)^{5/2}E_{51}^{-3/2}\,\mathrm{g\, s^{3}\, cm^{-3}}\label{eq2}\end{equation}
\begin{equation}
v_{e}=\left(\frac{E}{6M_{e}}\right)^{1/2}=2.44\times10^{8}\,\left(\frac{M_{e}}{\mch}\right)^{-1/2}E_{51}^{1/2}\,\mathrm{cm\, s^{-1}},\label{eq3}\end{equation}
where $\mch\approx1.4\, M_{\odot}$ and $E_{51}$ are, respectively, the Chandrasekhar mass and the
supernova energy in units of $10^{51}$~ergs.

We use the same scaling as \citet{DC98}:
\begin{equation}
R^{\prime}=\left(\frac{M_{e}}{4/3\pi\rho_{\mathrm{am}}}\right)^{1/3}\approx2.19\,\left(\frac{M_{e}}{\mch}\right)^{1/3}n_{0}^{-1/3}\,\mathrm{pc},\label{eq4}\end{equation}
\begin{equation}
V^{\prime}=\left(\frac{2E}{M_{e}}\right)^{1/2}\approx8.45\times10^{3}\,\left(\frac{E_{51}}{M_{e}/\mch}\right)^{1/2}\,\mathrm{km\, s}^{-1},\label{eq5}\end{equation}
and
\begin{equation}
T^{\prime}=\frac{R^{\prime}}{V^{\prime}}\approx248\,\left(\frac{M_{e}}{\mch}\right)^{5/6}E_{51}^{-1/2}n_{0}^{-1/3}\mathrm{\, yr},\label{eq6}\end{equation}
with $n_{0}=\rho_{\mathrm{am}}/(2.34\times10^{-24}$~g) being the number density of the interstellar
medium, assuming a 10:1 H:He ratio. Pressure values were scaled to the ram pressure
$\rho_{\mathrm{am}}V^{\prime2}$. This scaling is not unique; \citet{MT95} adopt a different scaling
in their treatment of SNRs. For this paper we adopt the convention that primed lowercase letters
denote the real, physical quantities, while unprimed lowercase letters are used for their scaled
counterparts, e.g. $t=t^{\prime}/T^{\prime}$.

Though \citet{DC98} considered multiple different ejecta profiles in their one-dimensional study,
we use only their exponential profile here. Other work \citep{FD10} has been performed in 3-D
using the power law profile, against which the results presented in this paper can be compared.
To our knowledge, no multi-dimensional simulations have used the constant profile.

\section{THE NUMERICAL MODEL}\label{sec:Numerical-models}

To run the simulations we used the hydrodynamics code VH-1, which solves the Euler equations for
fluid flow using the piecewise parabolic method with a Lagrangian remap step -- see \citet{CW84} for
a thorough description of the procedure. The base hydrodynamics code was extended with a subroutine
to calculate ionization age of ejecta elements, which could then be used to generate emission maps
for comparison against observations.

In order to save computing time, simulations were initiated in one dimension, then continued in
3-D at a later time. The starting time used in one dimension was one day, or
$t=2.74\times10^{-3}$~yr/$T^{\prime}=1.1\times10^{-5}(M_{e}/\mch)^{-5/6}E_{51}^{1/2}n_{0}^{1/3}$.
The grid at initialization started at $r_{\mathrm{min}}=5\times10^{-5}$ with
$r_{\mathrm{max}}/r_{\mathrm{min}}=2.4$, so that the forward edge of the grid was just beyond where
the exponential profile reaches a density equal to the ISM. The grid contained 480 zones, for a
resolution of $\Delta r/r_{\mathrm{max}}\approx1.2\times10^{-3}$; this was deemed to be an
acceptable balance between resolution and computing time for the three-dimensional runs (a brief
discussion of resolution effects occurs in the next paragraph, but all 3-D runs were performed at
this resolution). The initial conditions within the ejecta were a radial velocity of $v=r/t$,
density set according to the exponential profile, pressure calculated as an ideal gas at a
temperature of 5~K, and angular velocities set to 0. The ISM was assumed to be at rest, with a
density of unity, and cold. The interior radial boundary condition was set to the exponential model
with no angular velocity, and the exterior radial boundary matched the ISM described previously. To
follow the motion of the shocks over many doubling times, the grid tracked the forward shock,
advancing and expanding by a small amount whenever the forward shock moved within six zones of the
outer edge of the grid.

Early two-dimensional runs with an effective adiabatic index of $\gamma=5/3$ showed that resolution
had a minimal effect on the gross structure of the instabilities generated. At $t=2.0$ in
simulations with resolutions between $\Delta r/r_{\mathrm{max}}\approx4.9\times10^{-4}$ and
$\Delta r/r_{\mathrm{max}}\approx2.4\times10^{-3}$, the dominant wave mode was consistent across all
resolutions, though at higher resolutions more fine structure was present. Changing the resolution
has two effects beyond modifying the power spectrum of Rayleigh-Taylor instabilities. First,
increasing the resolution would allow for enhanced Kelvin-Helmholtz instability formation, which
should in turn accelerate the onset of the instability saturation period discussed in the next
paragraph. Second, as the effective adiabatic index decreases, a high resolution becomes necessary
to resolve the small-scale structure associated with highly compressible fluid.

One constant through all of the two-dimensional runs was the presence of three ``epochs'' regarding
instabilities. In the first of these, the initial seeds (random perturbations of either density,
radial velocity, or pressure; or a long-wavelength perturbation of radial velocity) generated
Rayleigh-Taylor fingers with a wavelength of a few grid zones, which then cascaded into larger
structures. The second epoch was a period of instability saturation: multiple cycles in which
Rayleigh-Taylor fingers appeared, grew toward the forward shock, experienced shear from the
Kelvin-Helmholtz instability along their edges, and fell back towards the reverse shock. An
instability ``freeze out'' divided the second and third epochs. After the freeze out, it became
possible to track individual RT structures to the end of the simulation, since production of new RT
structures slowed down and a second cascade to larger structures occurred. The instability
saturation period has been reported for both the power law \citep{CBE92,KDR99,BE01} and exponential
\citep{Dwar00,WC01} models. Because of this epoch, we expect that any trace of the initial
perturbation, at least for the small magnitudes applied in our simulations, is washed away before
the fingers freeze out.

We further investigated the saturation period by changing the times at which 1-D runs were expanded to two
dimensions and further evolved. One-dimensional runs were carried out to $t=3.65\times10^{-5}$ (the
earliest time at which the contact discontinuity appeared as a minimum in the density profile rather
than as a kink), $t=3.65\times10^{-4}$ and $t=3.65\times10^{-3}$ before mapping to two dimensions
and seeding instabilities. The 2-D run begun at $t=3.65\times10^{-4}$ achieved the saturation
exhibited by the earliest start time. The $t=3.65\times10^{-3}$ run never reached the instability
saturation epoch; the first generation of RT fingers were still identifiable at the end of the
simulation, at $t=2.0$. As long as the mapping from one to multiple dimensions occurred early
enough, the simulations achieved saturation and were largely indistinguishable. We therefore saved
computation time in the three-dimensional runs by starting them at $t=3.65\times10^{-4}$ without any
apparent changes to the evolution of the remnant. The importance of the instability saturation
period to our results bears restating: covering four or more decades of time allowed the
instabilities to saturate and wash out traces of the initial perturbations, while simulating three
decades resulted in clear imprints of the initial perturbations in the final state of the ejecta.
\citet{Orl12} used only two decades of expansion time in their simulations, from an age of 10 years
to an age of 1000 years, and their runs with smooth ejecta are still clearly dominated by grid
effects at late times (note the quadrilateral symmetry in figure~6 of that paper, a product of the
Cartesian grid used in the simulations).

Given the long-standing evidence in the literature that efficient cosmic-ray acceleration at shock
fronts can dramatically effect the eventual morphology of an SNR, three different three-dimensional
runs were performed. As an approximation for energy loss at shock fronts, we used the procedure of
\citet{BE01}, globally adjusting the adiabatic index of the fluid to increase its compressibility.
The run with an adiabatic index $\gamma=5/3$ (ideal gas with a strong shock compression ratio
$\sigma=4$) served as a control. To compare against this we performed simulations with
$\gamma=4/3$ ($\sigma=7$) and $\gamma=6/5$ ($\sigma=11$).

The three-dimensional runs covered the full $4\pi$ steradians, allowing instabilities everywhere in
the simulation to grow without boundary effects, and preventing grid-induced preferred wavenumbers. To
avoid geometrical singularities at the poles and allow for a more uniform $\Delta\theta$ and
$\Delta\phi$ across the grid, we employed specialized Yin-Yang grid of two congruent parts with
angular extent $\pi/2$ in $\theta$ and $3\pi/2$ in $\phi$ \citep{KS04}. A one-dimensional
kickstarting run was taken out to a scaled time of $t=3.65\times10^{-4}$, then swept across the grid
to form the basic three-dimensional profile, discarding the innermost quarter of the 1-D grid -- 
zones well inside the reverse shock even at the end of the simulation. The 3-D runs used a grid
resolution of $\Delta r/r_{\mathrm{max}}\approx1.2\times10^{-3}$, in keeping with the 1-D kickstart
runs. Though ``cubic'' zones with $\Delta r\approx r\Delta\theta\approx r\sin\theta\Delta\phi$ would
have been preferred, memory and computation time constraints required the angular resolution be
decreased by a factor of roughly two from the cubic value. The resolution of the three-dimensional
runs was ultimately $360\times360\times1080\times2$ ($r\times\theta\times\phi$, with the final
factor of 2 representing the Yin and Yang sections of the grid), with
$r_{\mathrm{max}}/r_{\mathrm{min}}=1.78$. We deemed resolutions lower than this to be insufficient
to capture the structures expected in the runs with lower adiabatic indices.

The instability seed for our simulations was a random multiplier between 0.95 and 1.05 applied
independently to the density of each cell containing shocked ejecta; two-dimensional runs showed
that the form of any initial perturbation has a negligible effect on the final shape of the
simulation remnant as long as the simulation is evolved long enough to include the instability
saturation period discussed in the previous paragraph.

Finally, ionization age of shocked ejecta was tracked to investigate its effect on observed
morphology. As used in our simulations, the ionization time is a computationally inexpensive
parameter that can then be used to provide a more accurate emission map than a basic density-squared
model can. We first assumed that the electron density $n_{e}$ is proportional to the fluid density
throughout the remnant. During each cycle we calculated the temperature of every ejecta element;
elements hotter than our cutoff temperature of $5\times10^{5}$~K were assumed to have been shocked,
and we updated the ionization time in each such cell by
$\tau_{\mathrm{new}}=\tau_{\mathrm{old}}+\rho\cdot dt$. Over the duration of our simulations,
shocked ejecta never dropped below the cutoff temperature, so we could track shocked ejecta solely
by temperature. We further assumed that iron is the sole contributor of free electrons to the
ejecta; this is a very rough approximation, and future work might include additional species of ions
for a more accurate picture. Using neon-like iron (16 free electrons per nucleus), the scale factor
for ionization age is
\begin{align}
\tau_{\mathrm{scale}} &= &&\frac{\rho_{\mathrm{am}}\cdot\mathrm{N_{\mathrm{A}}\cdot16}}{55.8}\cdot T^{\prime} \notag \\
&\approx &&3.15\times10^{9}\, n_{0}^{2/3}E_{51}^{-1/2}\left(\frac{M_{e}}{\mch}\right)^{5/6}\,\mathrm{cm}^{-3}\,\mathrm{s}.
\label{eq:7}\end{align}
where $\mathrm{N_{\mathrm{A}}}$ is Avogadro's number and 55.8 the molecular mass of iron.

\subsection{Dependence on time}\label{sub:Dependence-on-time}

As mentioned in section~\ref{sec:The-exponential-model}, the only parameter affecting the morphology
of a particular run is the scaled time. One of the primary questions addressed in this paper is whether
the fleecy ejecta structures observed in type Ia SNRs are consistent with homogeneous ejecta. If the
structures are solely due to hydrodynamic instabilities, their size and distribution could indicate
the age of the remnant, providing a constraint on the three key parameters of the exponential model
($M_{\mathrm{ej}}$, $n_{0}$, and $E_{51}$). For all three runs ($\gamma=5/3$, $\gamma=4/3$, and
$\gamma=6/5$) we tracked the simulations longitudinally in time.

\begin{figure*}
\centering
\includegraphics[height=75mm]{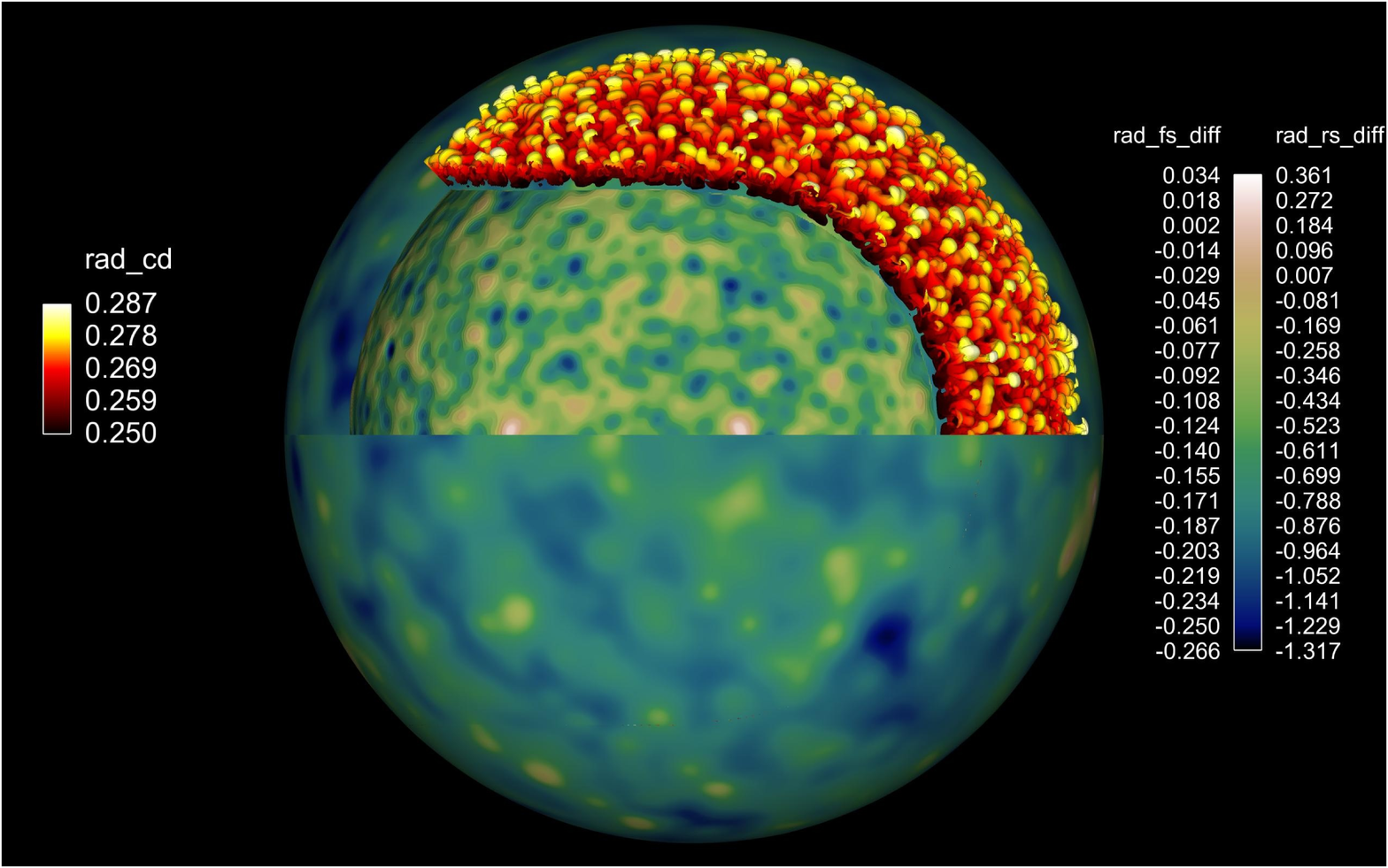} \\
\includegraphics[height=75mm]{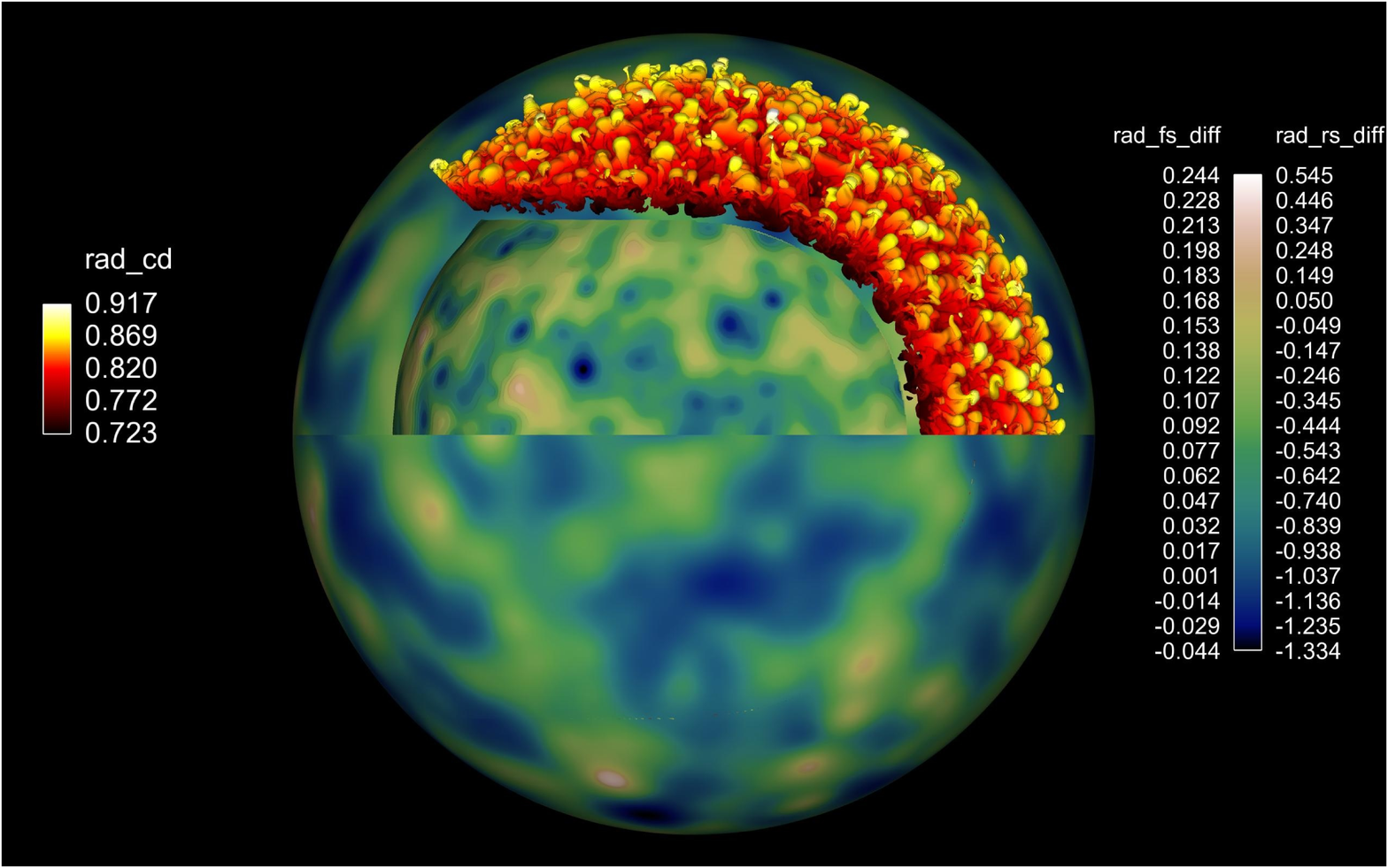} \\
\includegraphics[height=75mm]{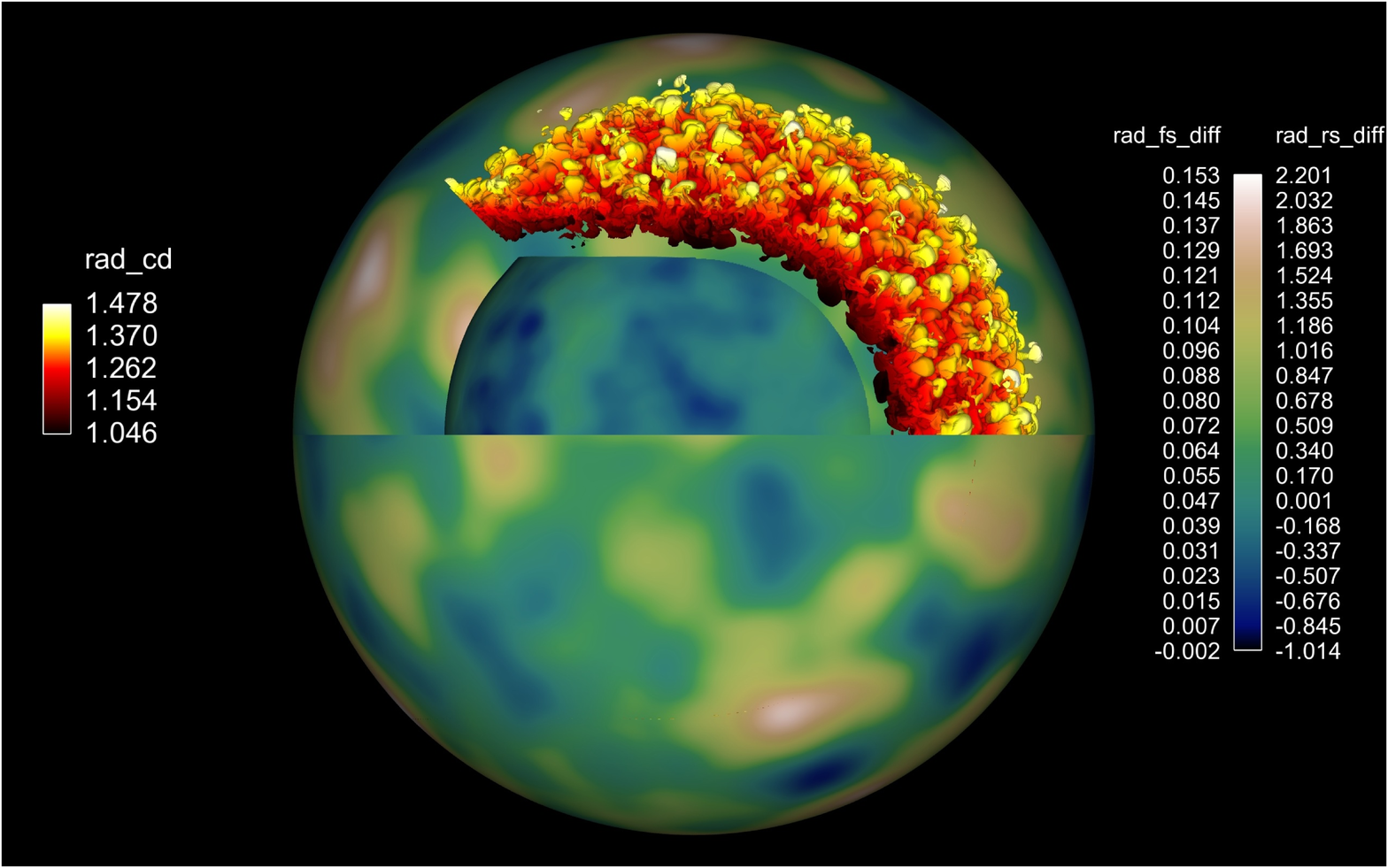}

\raggedright
\caption{\label{fig:3d-times}Isosurfaces for the reverse shock, contact discontinuity, and forward
shock for the $\gamma=5/3$ run at three times: $t=0.12$, $t=0.75$, $t=2.0$. The color scale for the
contact discontinuity is absolute radial position, while the color scale for the forward and reverse
shocks is percent difference from the average value for that interface as given in
table~\ref{tab:3d-interfaces-gam-53}.}
\end{figure*}
\begin{table}
\caption{\label{tab:3d-interfaces-gam-53}Interface Radii and Ratios from 3-D Data for $\gamma = 5/3$}
\begin{tabular}{cccccc} \hline
$t$ & $R_{\mathrm{RS}}$ & $R_{\mathrm{CD}}$ & $R_{\mathrm{FS}}$ & $R_{\mathrm{RS}}:R_{\mathrm{FS}}$ & $R_{\mathrm{CD}}:R_{\mathrm{FS}}$ \\
\hline
0.12 & 0.249 & 0.265 & 0.297 & 0.838 & 0.892 \\
0.75 & 0.716 & 0.800 & 0.952 & 0.752 & 0.840 \\
2.0 & 1.005 & 1.229 & 1.615 & 0.622 & 0.761 \\ \hline
\end{tabular}
Here and throughout the paper, $R_{\mathrm{CD}}$ is used to represent
$<R_{\mathrm{CD}}>$, the radius averaged over all $4\pi$ steradians.
\end{table}
\begin{figure}
\includegraphics[width=\columnwidth]{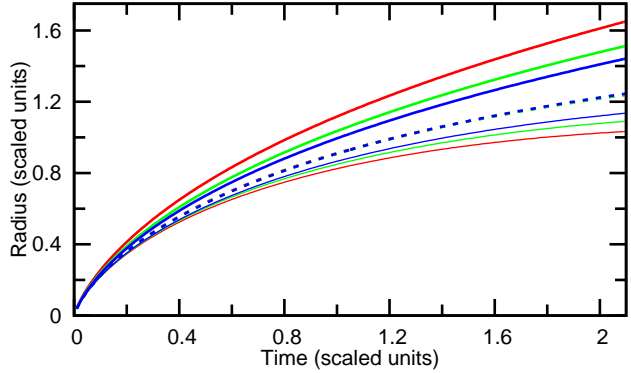}

\caption{\label{fig:1d-3d-compare-1}Interface locations as a function of time and adiabatic index
for the exponential model in one dimension. The forward shock is tracked by thick solid lines, the
reverse shock by thin solid lines, and the contact discontinuity by dashed lines. The curves for
$\gamma=5/3$, $\gamma=4/3$, and $\gamma=6/5$ are in red, green, and blue respectively. The three
curves for the contact discontinuity are separated by less than the width of the line used to show
them.}
\end{figure}

Figure~\ref{fig:3d-times} shows the $\gamma=5/3$ run at three different times. Isosurfaces have been
drawn of the three key interfaces -- the reverse shock, the contact discontinuity (where the ejecta
fraction of fluid elements is 0.5), and the forward shock. All images in figure~\ref{fig:3d-times}
are scaled to the same angular size, but the data presented in table~\ref{tab:3d-interfaces-gam-53}
provide the relative scale. All images occur after the instabilities freeze out, so the production
of new RT structures is negligible compared to the merging of existing structures. The three images
have several distinguishing features, such as the proximity of the various fluid discontinuities and
the structure at the forward and reverse shocks.

The three interfaces consistently diverge from each other over the course of the simulation,
illustrated by the one-dimensional simulations shown in figure~\ref{fig:1d-3d-compare-1}: the
contact discontinuity decelerates less than the reverse shock, and the forward shock decelerates
less than either. Consequently, as time progresses there is less interaction between the two shock
fronts and the Rayleigh-Taylor structures at the contact discontinuity. At the earliest time shown
in figure~\ref{fig:3d-times}, the interaction between the contact discontinuity and the reverse
shock is clear, as the features on the reverse shock are similar in angular size and spacing to the
Rayleigh-Taylor structures just outside them. As the reverse shock recedes from the contact
discontinuity, the shock front smooths out, the number of features and their radial extent shrinks,
and their angular size grows. A similar process can be seen occurring at the forward shock: by
$t=2.0$ the entire shock front has a radial dispersion of less than a single radial zone on the
grid.

\begin{figure}
\includegraphics[width=\columnwidth]{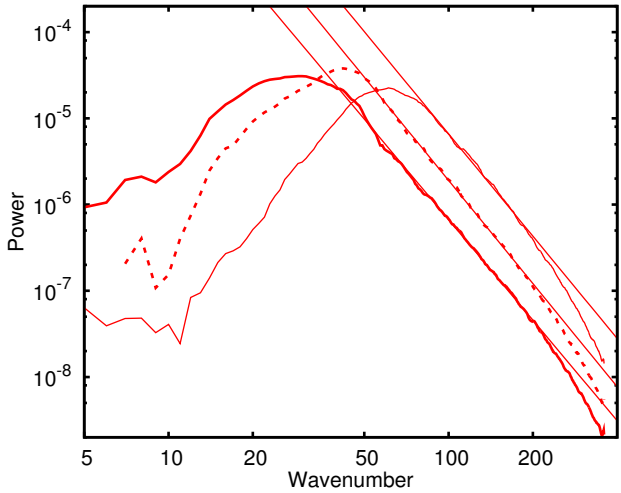}

\caption{\label{fig:temporal-53-2d}Power spectra for the radial ejecta column density in the
 $\gamma=5/3$ simulation. The curves for early time ($t\approx0.15$), middle time ($t\approx0.75$)
and late time ($t=2.0$) are thin solid line, dashed line, and thick solid line respectively. The
overlaid lines are best fits to power laws, $P\propto l^{-\alpha}$, with all exponents $\sim3.9$.
All spectra have been normalized and smoothed.}
\end{figure}

The images in figure~\ref{fig:3d-times} show that the Rayleigh-Taylor structures in shocked ejecta
trend toward larger angular size at later times. To further test this, we integrated the total
shocked ejecta density in radial columns over all $4\pi$ steradians. We then used the SHTOOLS
package\footnote{Available at http://shtools.ipgp.fr/} to calculate power spectra for the generated
image. The results are shown in figure~\ref{fig:temporal-53-2d}, which plots the normalized power
versus degree of spherical harmonic.  The power spectrum's peak moves to lower values of $l$ as the
simulation progresses, from $l\sim60$ at the earliest time to $l\sim30$ at $t=2.0$. The spectra
between the peak and $l\approx200$ can be fit to a power law in each case, with an exponent around
3.9. Higher than $l\approx200$ the bottleneck effect \citep{Dob03} caused a steepening of the power
spectra in every case shown. Though the slope of the power spectrum does not substantially change
with time, it is clear that power is transfered from higher wavenumbers to lower wavenumbers over
the course of the simulation.

All of the fits are significantly steeper than fitted power laws to the observed contact
discontinuity of Tycho's SNR, $\mathrm{P}\sim k^{-1.5}$ \citep{Warr05}. Further, the difference is
in the wrong direction: when the naturally two-dimensional surface of the CD is projected to a
single line around the remnant, projection effects should act to smooth out some of the
high-frequency power, steepening the power spectrum relative to that of the original 2-D surface.
(See Appendix A for additional information.) While both power laws extend to a
wavenumber of about 180, our peak occurs at a much higher wavenumber ($l\sim30$) than theirs
($l\sim6$).

\subsection{Dependence on adiabatic index}\label{sub:Dependence-on-gamma}

Support abounds in the literature for the idea that efficient cosmic-ray acceleration impacts the
evolution and morphology of SNRs. We find that increased shock compressibility leads to dramatic
differences in the morphology of the shocked ejecta and of the two shock fronts. In this section we
will describe the effect of the adiabatic index on four aspects of morphology: the shape and
location of the contact discontinuity, the shape and location of the forward and reverse shock
fronts, the power spectrum of the contact discontinuity, and the deceleration parameter.

\begin{figure}
\includegraphics[width=0.49\columnwidth]{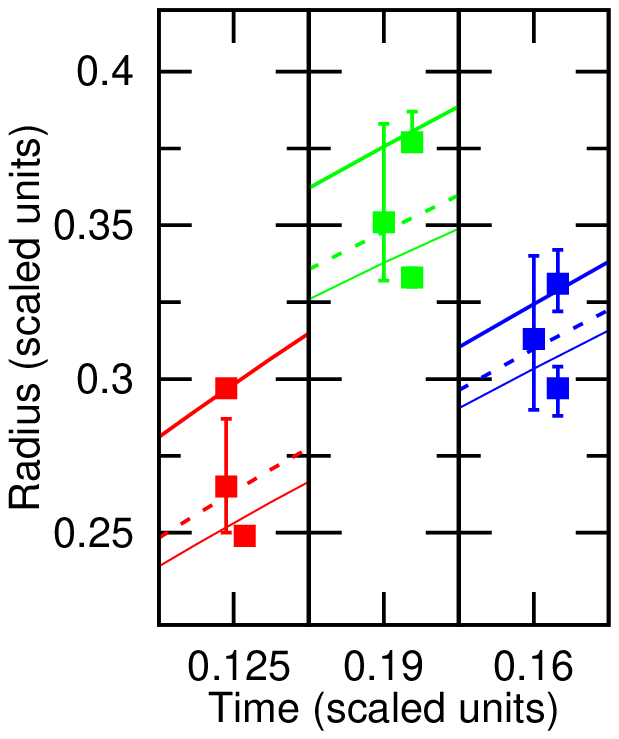}
\includegraphics[width=0.49\columnwidth]{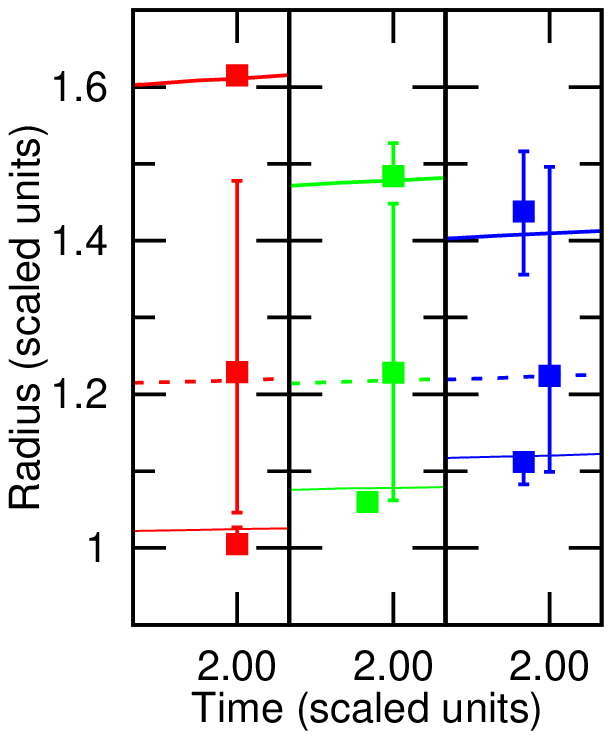}

\caption{\label{fig:1d-3d-compare-2}Interface locations as a function of time and adiabatic index
for the exponential model in one and three dimensions. Fluid discontinuities from 1-D runs are
colored and marked as in figure~\ref{fig:1d-3d-compare-1}. Interface locations in the 3-D runs are
listed in tables elsewhere in the paper. The plots for $\gamma=5/3$, $\gamma=4/3$, and $\gamma=6/5$
appear as the left, middle, and right panels in each triplet, respectively. \emph{Left}: Close-ups
of the 1-D interface locations around $t=0.15$, with the 3-D interface locations plotted for
comparison. The error bars show the maximum and minimum extent of the interface. To prevent overlap
of 3-D error bars, some shocks are plotted slightly to the right of their proper position.
\emph{Right}: As before, but around $t=2.0$; shifting here is to the left.}
\end{figure}
\begin{figure*}
\centering
\includegraphics[height=75mm]{fig1c.eps} \\
\includegraphics[height=75mm]{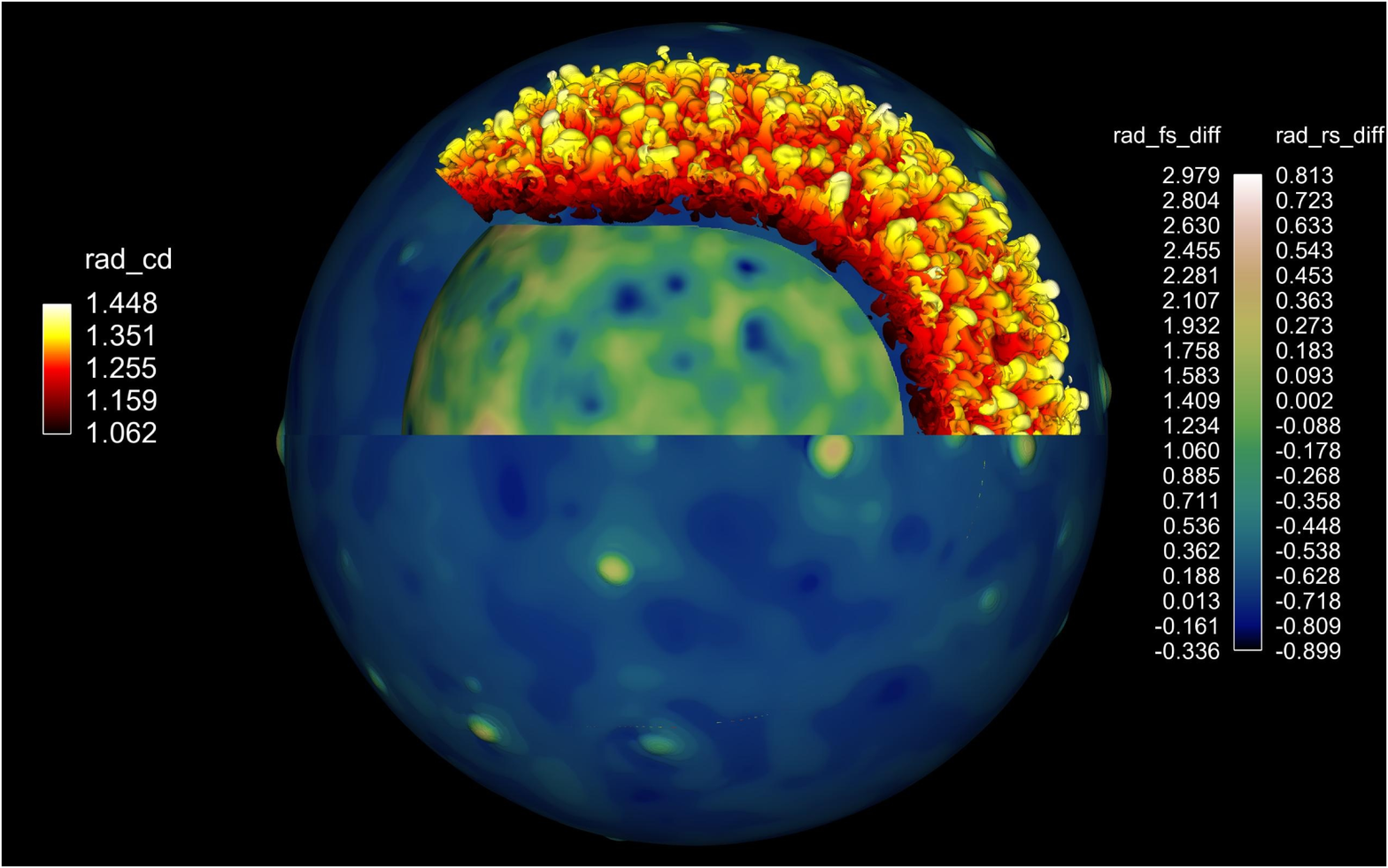} \\
\includegraphics[height=75mm]{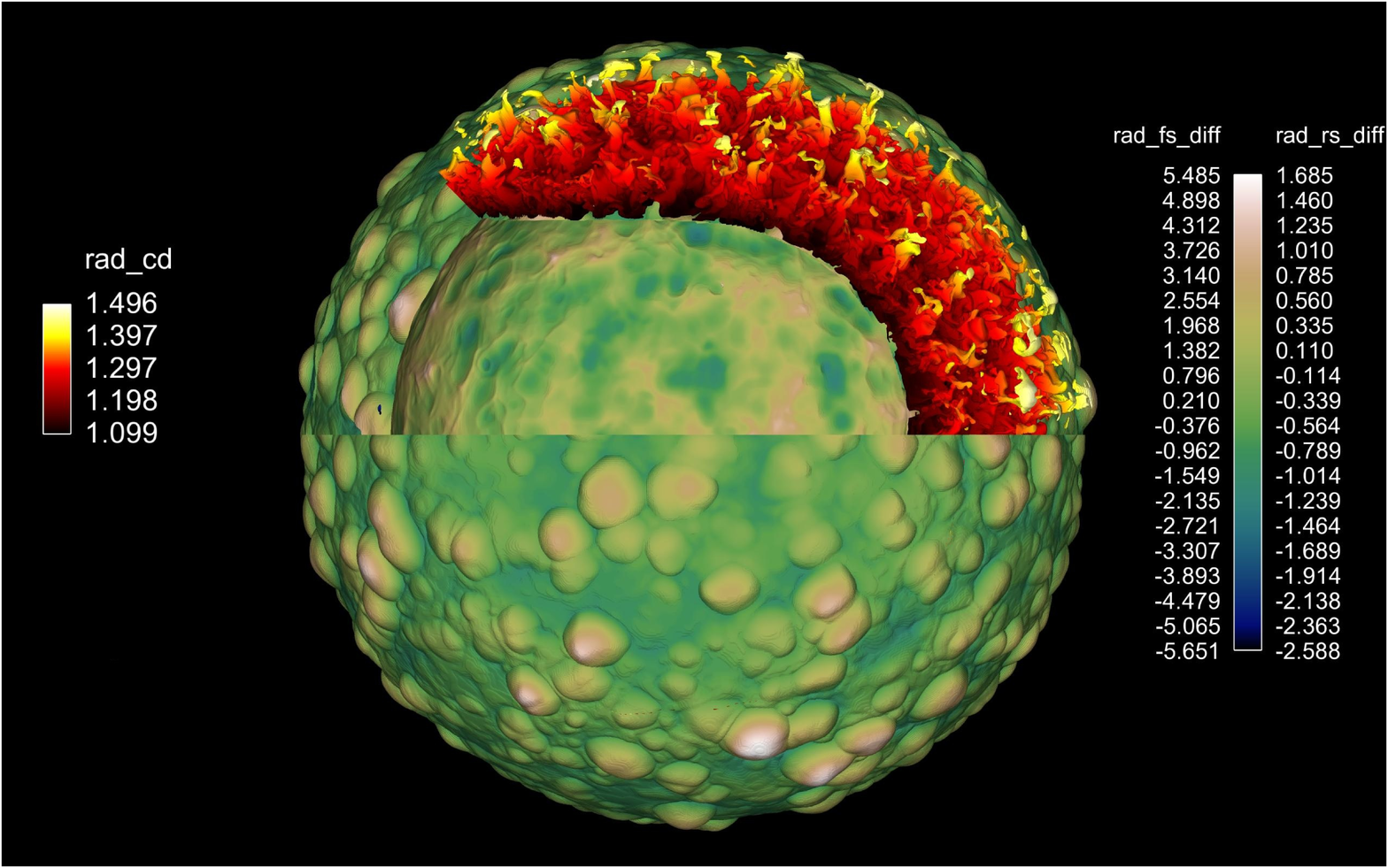}

\raggedright
\caption{\label{fig:3d-gammas}Isosurfaces for the reverse shock, contact discontinuity, and forward
shock at $t=2.0$ for all three runs. The color scale for the contact discontinuity is absolute
radial position, while the color scale for the forward and reverse shocks is percent difference
from the average value for that interface as given in table~\ref{tab:3d-interfaces-time-2.0}.}
\end{figure*}
\begin{table}
\caption{\label{tab:3d-interfaces-time-2.0}Interface Radii and Ratios from 3-D Data for $t = 2.0$}
\begin{tabular}{cccccc}
\hline
$\gamma$ & $R_{\mathrm{RS}}$ & $R_{\mathrm{CD}}$ & $R_{\mathrm{FS}}$ & $R_{\mathrm{RS}}:R_{\mathrm{FS}}$ & $R_{\mathrm{CD}}:R_{\mathrm{FS}}$ \\
\hline
$5/3$ & 1.005 & 1.229 & 1.615 & 0.622 & 0.761 \\
$4/3$ & 1.060 & 1.228 & 1.484 & 0.714 & 0.827 \\
$6/5$ & 1.112 & 1.224 & 1.438 & 0.773 & 0.851 \\ \hline
\end{tabular}
\end{table}

Under the assumption of spherical symmetry, reducing $\gamma$ from $5/3$ has the most immediate
morphological effect of changing the shock locations of the remnant at any particular time,
as illustrated in figure~\ref{fig:1d-3d-compare-1}. This does not apply to the contact
discontinuity: the difference between the three 1-D runs' contact discontinuities is less than the
width of the line used to plot their position in figure~\ref{fig:1d-3d-compare-1}. We explain this
with an appeal to conservation of momentum. In the thin shell limit of \citet{Gull73}, the entirety
of the shocked ejecta and the swept-up ISM are contained at a single radius. As the gas becomes less
compressible, both the ejecta piston and the shocked ISM regions expand in volume relative to the
thin shell limit. This means that both shocked ejecta and the shocked ISM increase in mass, sweeping
up additional matter that was beyond either shock front when the fluid was more compressible. The
additional inertia of the shell of ISM is roughly balanced out by the gained momentum of the piston,
and at a given time the radius of the contact discontinuity stays roughly constant as the
compression ratio drops from $\sigma=\infty$ to $\sigma=4$. This effect is not permanent: the
thickness of the regions of shocked fluid increases constantly and reduces the appropriateness of
the thin-shell approximation. After a few times $t^{\prime}$ (see section
\ref{sec:The-exponential-model}), runs with lower compressibilities start to decelerate relative to
runs with greater compressibilities. The likely ages of Tycho's SNR and 1006 fall well short of the
point of separation, however, so for the purposes of this work the location of the contact
discontinuity is independent of compressibility.

Beyond spherical symmetry, the shape of the interfaces can also vary, in some cases substantially.
Figure~\ref{fig:3d-gammas} shows all three three-dimensional runs at a scaled time of $t=2.0$. As
before, the remnants have been scaled to the same size, with the actual radii provided in
table~\ref{tab:3d-interfaces-time-2.0}.

In all of the tables in this paper, we list the location of the contact discontinuity as a single
number, its average location. In actuality the contact discontinuity has a very complex shape that
depends on both the remnant's age and the compressibility of the fluid, as can be seen in
figures~\ref{fig:3d-times} and \ref{fig:3d-gammas}. By $t=2.0$, the CD is spread out over many
radial zones (its radial extent is as high as 29\% of its maximum radius for the images in
figure~\ref{fig:3d-gammas}), and at many places around the remnant occurs multiple times in a single
radial column (e.g., locations with RT mushroom caps above the base of the structure).  Using just
the average location of the contact discontinuity is a great simplification. Reducing the structure
of the CD to a single number is nonetheless justified: as in one dimension (see
figure~\ref{fig:1d-3d-compare-1}), table~\ref{tab:3d-interfaces-time-2.0} shows that the average
radius of the contact discontinuity in three dimensions is nearly constant across all three values
of $\gamma$ at $t=2.0$. Despite the independence of the CD's average radius and the compressibility
of the gas, as $\gamma$ decreases there is a bias towards more ejecta close to the reverse shock;
this is visible in figure~\ref{fig:3d-gammas} as a shift in color of the contact discontinuities
away from yellow/white and toward red/black.

The average location of the forward and reverse shocks in each of the 3-D runs is slightly less
consistent with their radial positions in the corresponding 1-D simulations, as noted in
figure~\ref{fig:1d-3d-compare-2} (the forward and reverse shocks have been plotted slightly to the
left or right when needed for the sake of clarity). Both shock fronts are slightly more advanced in
three dimensions -- the reverse shocks are at a lower average radius, and the forward shocks at a
greater average radius. The advanced location of the forward shocks is likely due to interaction
between the shock front and Rayleigh-Taylor fingers at the contact discontinuity. As the adiabatic
index decreases and interaction between the two interfaces increases, protrusions appear at the
forward shock that pull the average location ahead of where it would be in one dimension.
Figure~\ref{fig:1d-3d-compare-2} also shows error bars marking the maximum and minimum radii for
each interface, further demonstrating the interaction and bubbles already mentioned. 

Figure~\ref{fig:3d-gammas} offers clear support for the effect of Rayleigh-Taylor fingers on the
shape and location of the forward and reverse shocks. The $\gamma=5/3$ run shows very smooth forward
and reverse shocks -- the forward shock is located in the same radial zone everywhere in the remnant,
while the reverse shock is spread over just a few zones. There is a large gap between both shocks
and the contact discontinuity, the likely reason for the smoothness of both shock fronts. Although a
major assumption behind our simulations was smooth ejecta, the second and third images in
figure~\ref{fig:3d-gammas} demonstrate that smooth ejecta alone is not sufficient to guarantee
smooth forward and reverse shocks. The $\gamma=4/3$ run shows definite evidence of interaction at
the forward shock: the majority of the interface is as smooth as its $\gamma=5/3$ counterpart, but a
few tens of bumps can be seen where Rayleigh-Taylor fingers reached far enough outward to perturb
the sphericity of the forward shock. The situation is repeated at the reverse shock, where features
on the same angular scale as the RT structures are visible and the shock front itself has a radial
extent of $\approx0.02$ times its average radius in the data. When $\gamma=6/5$, the increased
compressibility of the fluid has a marked effect on the shape of the remnant. There is abundant
evidence of interaction between the contact discontinuity and the forward and reverse shocks (at the
right edge of the image a Rayleigh-Taylor finger can even be seen in the process of creating one of
the numerous bubbles visible on the forward shock).

\begin{figure}
\includegraphics[width=\columnwidth]{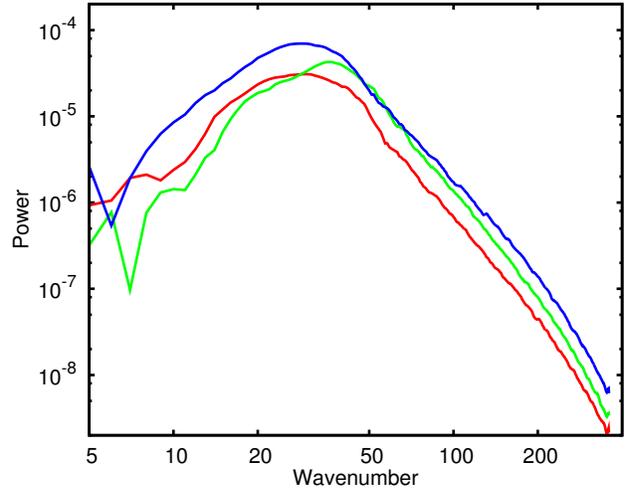}

\caption{\label{fig:gamma-t-late-2d}Normalized power spectra for the radial ejecta column density
of the 3-D simulations at $t=2.0$. The curves for $\gamma=5/3$, $\gamma=4/3$, and $\gamma=6/5$ are
in red, green, and blue respectively. The exponents to fitted power laws (not plotted here) are 3.9
for $\gamma=5/3$, 4.0 for $\gamma=4/3$, and 3.5 for $\gamma=6/5$.}
\end{figure}

As the adiabatic index decreases from 5/3 to 1, the ejecta become more compressible and
Rayleigh-Taylor structures can be thinner. This results in more power at small wavelengths in the
simulations with $\gamma<5/3$, illustrated in figure~\ref{fig:gamma-t-late-2d}, which was created in
the same manner as figure~\ref{fig:temporal-53-2d}. The figure shows an essentially monotonic
increase in power at high wavenumbers as $\gamma$ decreases (the $\gamma=4/3$ run's power spectrum
peaks later, around $l=40$, than do the power spectra of the other two runs, which peak around
$l=30$). Additionally, fitted power laws become shallower as the adiabatic index approaches 1, with
the exponent dropping from 3.9 ($\gamma=5/3$) to 3.5 ($\gamma=6/5$).

\begin{figure}
\includegraphics[width=\columnwidth]{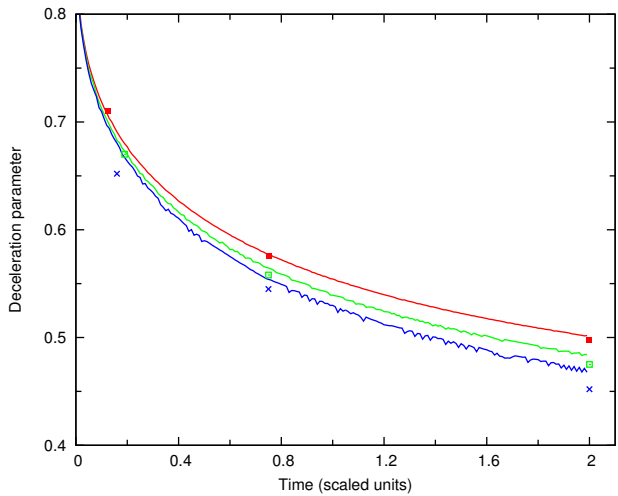}

\caption{\label{fig:1d-3d-decel-param}The deceleration parameter as a function of time and adiabatic
index for the exponential model. The curves for one dimension and $\gamma=5/3$, $\gamma=4/3$, or
$\gamma=6/5$ are in red, green, or blue respectively. Overlaid are points showing the calculated
deceleration parameter for the 3-D runs at selected times.}
\end{figure}

The deceleration parameter ($m=vt/r$) of the forward shock, as shown in
figure~\ref{fig:1d-3d-decel-param}, is another point of comparison between the radial structure in
1-D and its averaged equivalent in 3-D. Calculation of $m$ in the 1-D case is simple, as one can
easily track the motion of the forward shock, to grid-zone accuracy in space and arbitrary accuracy
in time, and arrive at a value for the forward shock expansion velocity. For the 3-D runs, we used
the Rankine-Hugoniot conditions for a strong shock to calculate the forward shock velocity in terms
of the downstream pressure, upstream density, and adiabatic index. The various deceleration
parameters are collected in figure~\ref{fig:1d-3d-decel-param}. As in
figure~\ref{fig:1d-3d-compare-2}, the deceleration parameter in one dimension for each value of
$\gamma$ is shown as a curve, with the corresponding 3-D values overlaid as points. At any
particular time decreasing the adiabatic index reduces the deceleration parameter, and a lowered
adiabatic index reduces the time at which the remnant reaches a particular value of the deceleration
parameter. The curves and points of figure~\ref{fig:1d-3d-decel-param} are in close agreement
everywhere, with the exception of the $\gamma=6/5$ run. There, the bubbles at the forward shock
lead to a substantial solid angle where the shock normal isn't radial. This induced angle appears as
a smaller radial expansion speed $v$, and so a smaller deceleration parameter. Restricting the
averaging process to only those grid zones on the forward shock whose normal is within $15^{\circ}$
of the radial direction eliminates most of the bubbles but retains the largely spherical base
visible in figure~\ref{fig:3d-gammas}. It also moves the calculated 3-D deceleration parameters to
within a few percent of the 1-D values, in line with the other values for $\gamma$; the corrected
values are shown in figure~\ref{fig:1d-3d-decel-param}, rather than the uncorrected numbers.

\section{OBSERVATIONAL IMPLICATIONS}\label{sec:Observational-implications}

To facilitate comparison with the X-ray observations of Tycho and SN 1006, we now present
line-of-sight projections from the three-dimensional data sets. We assume thermal emission from
shocked ejecta to be proportional to the square of gas density. To mimic the effects of emission
turn-on as shocked ejecta are ionized, a cutoff in ionization age is implemented by which emission
from some ejecta elements can be excluded (see section~\ref{sub:ionization-age-cutoffs}).
Synchrotron emission from the forward shock is visualized assuming that both relativistic electron
energy density $u_{e}$ and magnetic field energy density $u_{B}$ are proportional to the pressure
(i.e. nonlinear amplification of the magnetic field). Then, since synchrotron volume emissivity
$j_{\nu}$ is proportional to $u_{e}B^{(s+1)/2}$, with $s$ being the electron energy index (i.e.
$N(E) \propto E^{-s}$) \citep{Pach70}, we have $j_{\nu} \propto P^{(s+5)/4}$. As $s \cong 2.2$ for
both SN 1006 and Tycho \citep{Green09}, the synchrotron emissivity is calculated as $P^{1.8}$. X-ray
synchrotron emission decays more rapidly away from the forward shock than radio emission, so our
crude model results in a more diffuse shell of emission around the ejecta than would be present with
a more refined treatment. In all images of projections in this section (with the exception of
figure~\ref{fig:gamma-65-knots}), ejecta emission is in white and synchrotron emission is in purple.

Using these images, we take a more in-depth look at the distribution and character of emission from
shocked ejecta, the shape and location of the projected contact discontinuity, and the strength of
the limb brightening effect under ionization age cutoffs to emission.

\subsection{Fleece}\label{sub:Fleece}

\begin{figure}
\includegraphics[width=\columnwidth]{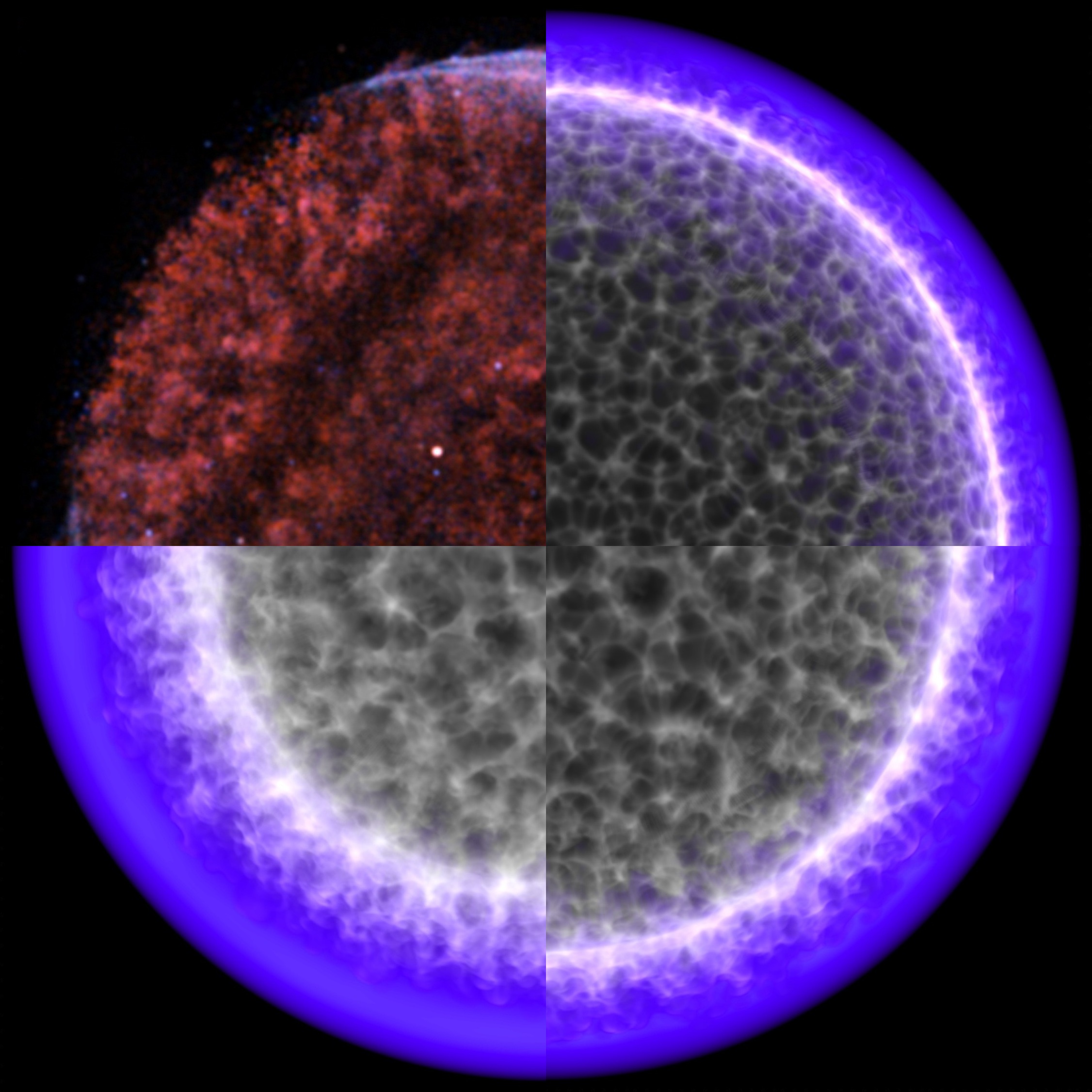}

\caption{\label{fig:sn1006-times}Images showing the effect that dynamical age has on the morphology
of the remnant. The SE quadrant of SN 1006 is included at the top left for comparison. All three
remaining images come from the $\gamma=5/3$ run. Clockwise from the top right: $t=0.12$, $t=0.75$,
$t=2.0$. Image of SN 1006 taken from \citet{CC08}.}
\end{figure}
\begin{figure}
\includegraphics[width=\columnwidth]{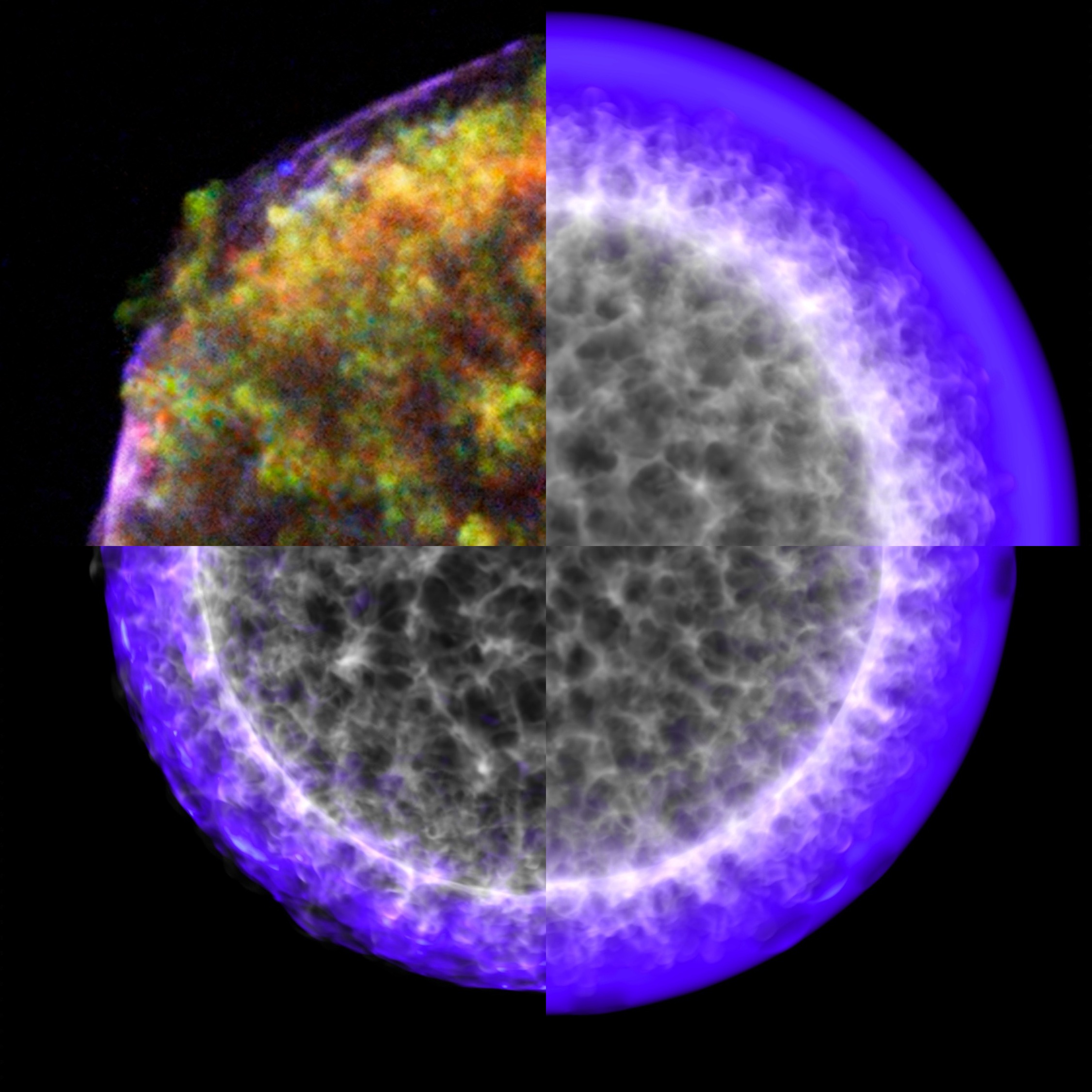}

\caption{\label{fig:tycho-gammas}Images showing the effect that changing the adiabatic index
$\gamma$ of the simulation has on the resultant remnant. The SW quadrant of Tycho's SNR is included
at the top left for comparison. Clockwise from the top right: the $\gamma=5/3$ run, the $\gamma=4/3$
run, and the $\gamma=6/5$ run. All three projections are scaled to the correct relative size so
interface locations can be directly compared. Image of Tycho taken from \citet{Warr05}.}
\end{figure}

Figure~\ref{fig:sn1006-times} compares the SE quadrant of SN 1006 against projected quadrants from
the $\gamma=5/3$ run at the same three times shown (in 3-D) in figure~\ref{fig:3d-times}.
Immediately apparent is that the emission from shocked ejecta looks very similar to the fleecy
complexes detected in X-rays in both Tycho and SN 1006. The earliest time shown in
figure~\ref{fig:sn1006-times} (top right) is just after the end of the instability saturation period
mentioned in section~\ref{sec:Numerical-models}. Ejecta structures at this time in the central
region are visible as a filamentary network. In the time between the early image and the end of the
simulation (bottom left of the figure) the Rayleigh-Taylor instabilities have grown, sheared, and
merged into each other. Figure~\ref{fig:tycho-gammas} compares the three different runs at $t=2.0$
to the SW quadrant of Tycho's SNR to illustrate the effect of compressibility on remnant morphology.
The rounded mushroom caps of the Rayleigh-Taylor structures at $\gamma=5/3$ are still noticeable at
$\gamma=4/3$, though the center of the remnant is darker relative to the limb brightening. By
$\gamma=6/5$ the RT structures' caps are no longer generally round: the greater compressibility has
resulted in much longer, thinner fingers of shocked ejecta, and extensive interaction with the
forward shock has bent or otherwise warped many of them.

\subsection{CD shape, relation to FS}\label{sub:shock-locations}

Figure~\ref{fig:sn1006-times} compared the $\gamma=5/3$ remnant at three different times to one
quadrant of SN 1006. The structures visible at the center of SN 1006 are smaller in angular size
than the RT structures at the center of the $t=2.0$ image, but also less filamentary than the
similarly-located ejecta at $t=0.12$. The RT fingers in both the $\gamma=5/3$ and the $\gamma=4/3$
remnant are largely oriented along radial lines, as seem to be the structures at the edge of SN
1006. Furthermore, the structures at the edge of SN 1006 appear to be more discrete, that is, with a
better-defined edge to each structure. The enhanced edges visible in the structures of SN 1006 allow
them to stand out against each other, as opposed to the less distinct haze outside the limb
brightening in the $t=0.75$ and $t=2.0$ images. There is no obvious cutoff in limb brightening to
mark a reverse shock in the image of SN 1006, as the contrast is dominated by the large-scale gap in
emission in the SE quadrant.

As the compressibility of the fluid increases, interaction with the contact discontinuity can cause
bubbles at the forward shock, noticeable in both the $\gamma=4/3$ and the $\gamma=6/5$ images in
figure~\ref{fig:3d-gammas}. However, these bubbles are much fainter at their maximum radial extent
than the shocked ISM at their base, so they do not generally show up in projection. One such bubble
is visible in the $\gamma=4/3$ quadrant of figure~\ref{fig:tycho-gammas} as a thin bright rim of
emission ahead of a darker patch. The situation is more extreme still with the $\gamma=6/5$ run,
where the bubbles comprise a much larger fraction of the forward shock. Instead of the mostly smooth
emission of the $\gamma=5/3$ and $\gamma=4/3$ runs, the projected forward shock of the $\gamma=6/5$
run shows up as a chaotic network of projected bubbles, with single bubbles at the edge of the
remnant too dim to appear in projection. It is for this run that the approximation to X-ray
synchrotron emission (discussed at the start of the section) is most telling; if the emission
decayed more quickly, the diffuse shell of emission would resolve into a filamentary network tracing
out the locations of the bubbles at the forward shock. Comparison between
figures~\ref{fig:3d-gammas} and \ref{fig:tycho-gammas} implies that the locations around the rim of
Tycho's SNR where emission from the forward shock is absent could be artifacts of interaction
between the shocked ejecta and the forward shock.

\begin{figure}
\includegraphics[width=0.49\columnwidth]{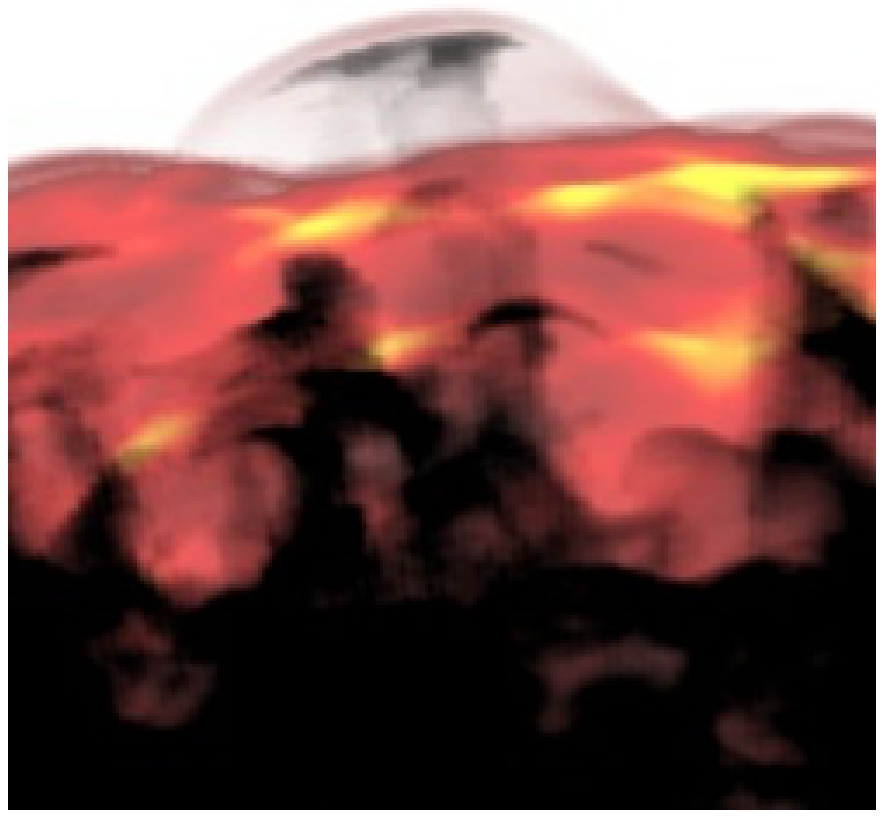}
\includegraphics[width=0.49\columnwidth]{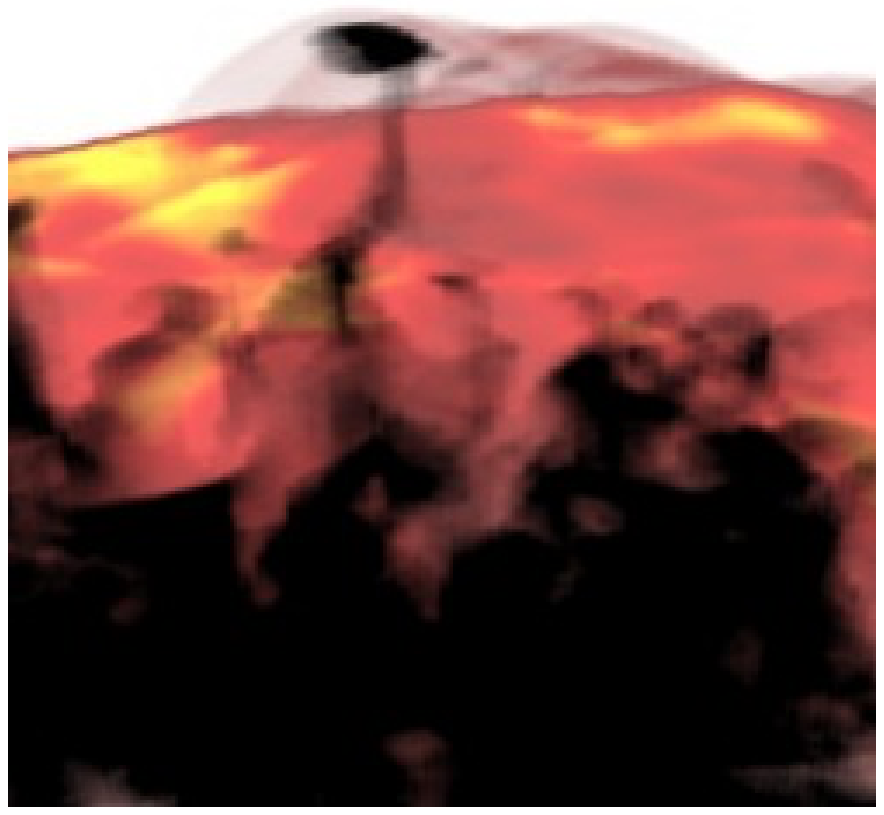}

\caption{\label{fig:gamma-65-knots}Magnified views of two locations from the $\gamma=6/5$ run where
knots of ejecta seem to have overtaken the forward shock.}
\end{figure}

One feature captured by the $\gamma=6/5$ run, not present in the other two, is locations around the
edge where knots of ejecta seem to be visible beyond the forward shock. Two examples have been
enlarged and recolored for ease of viewing in figure~\ref{fig:gamma-65-knots}. The effect is an
illusion caused by the faint bubbles of forward shock emission discussed in the previous paragraph,
and by projection effects; the contact discontinuity is always inside the forward shock in the
three-dimensional data. Even in the $\gamma=6/5$ run, such protrusions are rare, occurring only a
few times around the rim of the remnant; they are absent entirely from the $\gamma=4/3$ and
$\gamma=5/3$ runs. With the smooth ejecta used in our runs, the ISM must be highly compressible
before the forward shock is close enough to the contact discontinuity to be significantly affected
by RT fingers. An alternative explanation was suggested by \citet{Orl12}, who conclude that the
separation between the CD and FS is an indication of ejecta structure rather than of cosmic ray
acceleration: overdensities in the ejecta drive instability growth, resulting in more interaction at
higher compressibilities than seen in our simulations.

The knots pictured in figure~\ref{fig:gamma-65-knots} are roughly the same angular size,
$\lesssim10^{\circ}$, as the protrusions around the rim of SN 1006 (in the SE, S, and SW) and Tycho
(in the S and W). We conclude that these features can be generated by fluid instabilities alone,
without any inhomogeneities present within the unshocked ejecta or the unshocked ISM. There are no
features in the $\gamma=6/5$ run on the same scale as the shelf of thermal emission at the N rim of
Tycho or the tiered structure in the NE polar cap of SN 1006, both of which are many tens of degrees
across. The inability of the Rayleigh-Taylor instability to produce these features points at
inhomogeneities in the ejecta, the magnetic field strength around the progenitor, or the ISM.

\subsection{Ionization age cutoffs, limb brightening, and angular variation}\label{sub:ionization-age-cutoffs}

\begin{figure}
\includegraphics[width=\columnwidth]{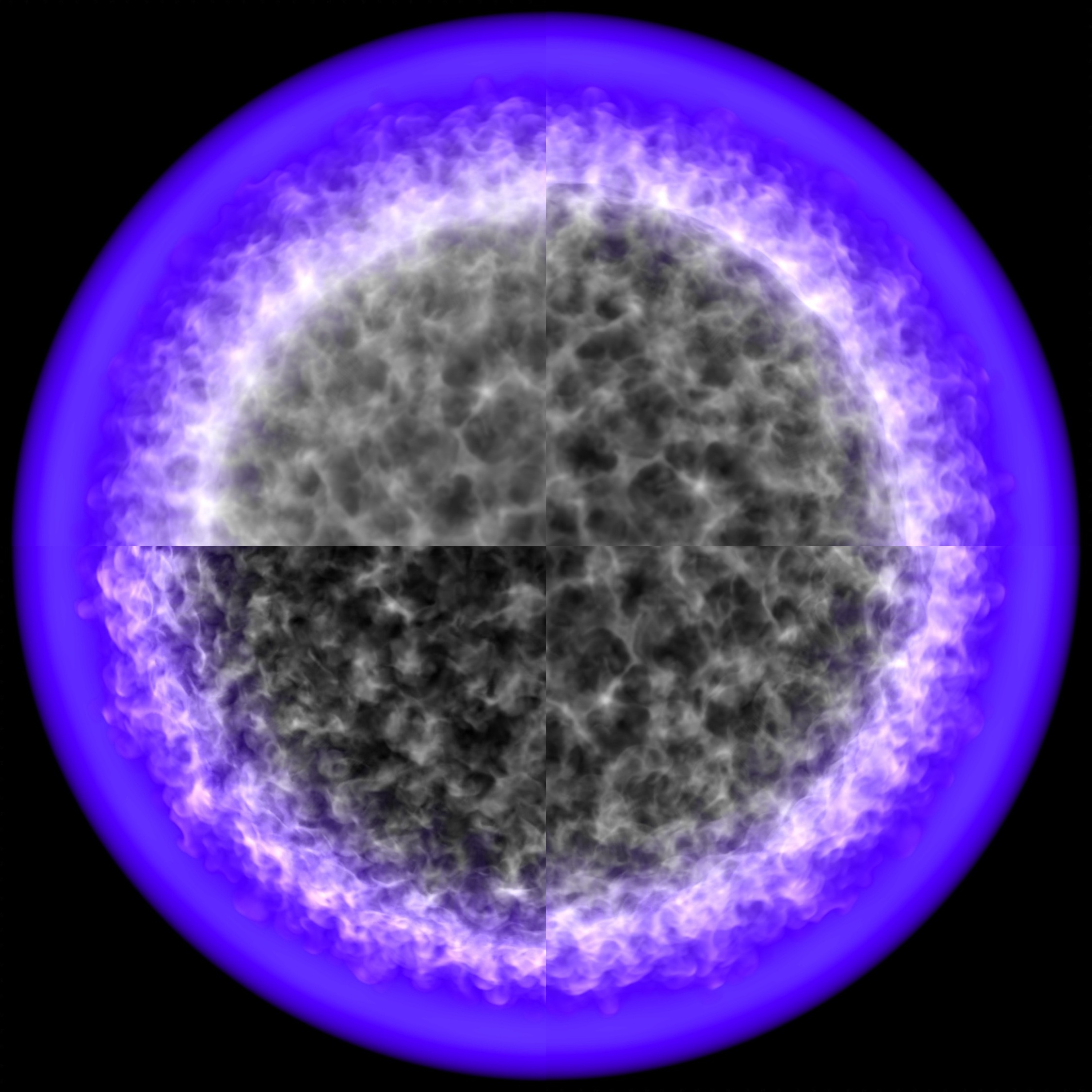}

\caption{\label{fig:ion-cutoffs}Above, images of one simulated remnant with increasingly high
cutoffs. \emph{Top left}: an image with no cutoff in place; \emph{top right}:
$\tau_{\mathrm{min}}=0.8$; \emph{bottom right}: $\tau_{\mathrm{min}}=1.6$; \emph{bottom left}:
$\tau_{\mathrm{min}}=2.4$. Using the scale factor from equation \ref{eq:7} and setting all
parameters to unity, the three cutoffs correspond in physical units to
$\tau_{\mathrm{min}}=2.52\times10^{9}\,\mathrm{cm}^{-3}\,\mathrm{s}$,
$\tau_{\mathrm{min}}=5.04\times10^{9}\,\mathrm{cm}^{-3}\,\mathrm{s}$,
and $\tau_{\mathrm{min}}=7.56\times10^{9}\,\mathrm{cm}^{-3}\,\mathrm{s}$ respectively. All images
are at $t=2.0$ for the $\gamma=5/3$ run.}
\end{figure}

Assuming that X-ray thermal emissivity is proportional to $n^{2}$ in visualizing data potentially
overestimates emission from the remnant: not all matter is radiating in all wavelengths at all
times, due at least in part to deviations from ionization equilibrium. To account for this effect in
the three-dimensional simulations, ionization age cutoffs were implemented during visualization,
below which shocked ejecta were assumed to be X-ray faint. The effect on observed morphology is
shown in figure~\ref{fig:ion-cutoffs}, which (clockwise from the top left) sets the cutoff for
emission at successively higher levels for the $\gamma=5/3$ run at $t=2.0$. There is no cutoff for
the top left quadrant ($\tau_{\mathrm{min}}=0.0$), and by the bottom left a high cutoff
($\tau_{\mathrm{min}}=2.4$) has eliminated about half of the shocked ejecta by volume. Since
recently shocked ejecta is (for the $\gamma=5/3$ run) well within the innermost extent of the
contact discontinuity, the cutoff preferentially eliminates smooth shells of recently shocked ejecta
at low $\tau$ and leads to raggedness at the inner edge of emission.

\begin{figure}
\includegraphics[width=\columnwidth]{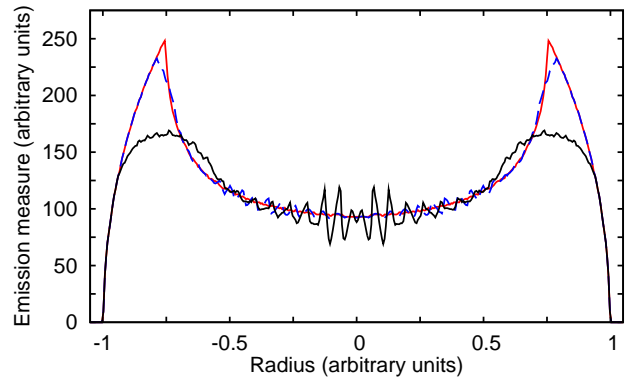}

\caption{\label{fig:projection-effects}Illustration of the effect of a disrupted inner edge on the
emission from a hollow shell. The solid red line shows a projection through a 2-D hollow shell; the
radius of the inner edge is 0.75 times that of the outer edge. The dashed blue line is a projection
through the same shell, but with the inner surface perturbed as $1+0.05\sin(40\theta)$. The solid
black line is similar, but with the perturbation given by $1+0.25\sin(40\theta)$. Towards the center
of the image the perturbations are nearly parallel to the line of projection, accounting for the
noise visible in the two perturbed cases.}
\end{figure}

This disruption of the inner boundary to emission primarily alters the strength of the limb
brightening effect. This outcome is illustrated in figure~\ref{fig:projection-effects}, in which we
generated two dimensional uniform hollow shells, perturbed the interior edge, then projected them
from two dimensions to one. Without any perturbations in place, the inner edge is clearly
identifiable as the sharp peak in limb brightening; in projections from three dimensions to two, a
pronounced drop in brightness would occur. With a regular perturbation of amplitude 5\% to the inner
radius, the peak of emission no longer exactly traces the unperturbed reverse shock, even though the
average radius is unchanged. Instead, the peak is outside that of the unperturbed case, and while
the peak of emission is still plain the decay towards the center is no longer as pronounced. When
perturbations are 25\% of the average radius (an extreme case that isn't suggested by the
three-dimensional data), the maximum of emission is now around that of the unperturbed case, but it
is no longer a clear peak. Both the interior and exterior regions to the radius of maximum emission
show gentle slopes, and the maximum itself is only $\sim150\%$ of the emission from near the axis of
symmetry of the shell.

There is definite limb brightening occuring in Tycho's SNR, but the inner edge to the emission is
indistinct. This effect is reproduced successfully by the lower two quadrants in
figure~\ref{fig:ion-cutoffs}, although the upper two quadrants (with no cutoff or a low one) still
display the reverse shock as a localized drop in emission. Comparison to SN 1006 is less favorable;
the limb brightening effect is either across too thin a region (i.e. the northwest quadrant) or too
broad a region (the southeast quadrant).

\section{DISCUSSION}\label{sec:Discussion}

This section elaborates on three topics that have been mentioned only in passing so far. We examine
the location of the interfaces (reverse shock, contact discontinuity, and forward shock) in the
projected images of figures~\ref{fig:sn1006-times} and \ref{fig:tycho-gammas}. We then combine
available information throughout the paper with observed quantities to estimate the dynamical age of
both SN 1006 and Tycho's SNR. Lastly, we discuss projection effects and associated error with an eye
toward future observations of remnants.

\subsection{Interface locations}

Addressing whether our remnants generate the radial structure of Tycho's SNR and SN 1006 requires an
analysis of our remnants mimicking the method of \citet{Warr05} for locating interfaces. With only
one ``component'' representing ejecta instead of nearly a dozen, however, we cannot replicate their
procedure exactly. We first divided the remnant into 1440 angular wedges $0.25^{\circ}$ in width,
and created a radial grid with 240 points along each angular wedge. Matching our remnant's radius
to the 251$^{\prime\prime}$ of Tycho's SNR, our radial resolution is roughly 1$^{\prime\prime}$.
This is finer than the value of 3$^{\prime\prime}$ used in \citet{Warr05}, but the resolution of our
data is sufficient to allow it: $\Delta r/r=1.2\times10^{-3}$, so the angular size of a single zone
at the edge of the projection is just 0.3$^{\prime\prime}$.

With the projection partitioned into a grid, we located the contact discontinuity, forward shock,
and reverse shock at each angular location. We identified the contact discontinuity with the
outermost radius where emission from shocked ejecta occurred. This method does not track the true
contact discontinuity (shown in figures~\ref{fig:3d-times} and \ref{fig:3d-gammas}): as discussed in
Appendix~\ref{app:proj-effects}, it biased toward protrusions at greater radius than the average
value across the remnant. It is nonetheless very similar to the identification process used to find
the CD in \citet{Warr05} and \citet{CC08}. The forward shock was defined as the outermost radius at
which intensity reached half the maximum value for each radial spoke. This misses some of the
filamentary structures that appear at lower values of $\gamma$, but does approximate the method of
\citet{Warr05}.

The reverse shock was found by treating the shocked ejecta as a hollow shell and identifying the
maximum value along each radial line. This correctly identifies the location of the reverse shock
for every combination of $\gamma$ and $t$ except $\gamma=5/3$, $t=2.0$. In that instance, the
density gradient in the shocked ejecta causes stronger emission near the contact discontinuity than
near the reverse shock. This is visible in the upper left quadrant of figure~\ref{fig:ion-cutoffs}
as a bright ring inside the edge of ejecta emission. The inner edge of emission becomes increasingly
ragged or diffuse as $\gamma$ decreases or as we exclude freshly-shocked ejecta from emission
calculations (see section~\ref{sub:ionization-age-cutoffs} above). Any maximum value less than 50\%
of the overall maximum value was deemed to be unresolvable against the background of Rayleigh-Taylor
fingers. Because of the Yin-Yang grid used for the simulations, overlap between the two parts of the
grid could generate an artificially high value for projected emission at the two angles where the
overlap occurs. To prevent this artifact from affecting the 50\% threshhold of the reverse shock,
the single highest value over all 1440 angular wedges was excluded.

\begin{figure}
\includegraphics[width=\columnwidth]{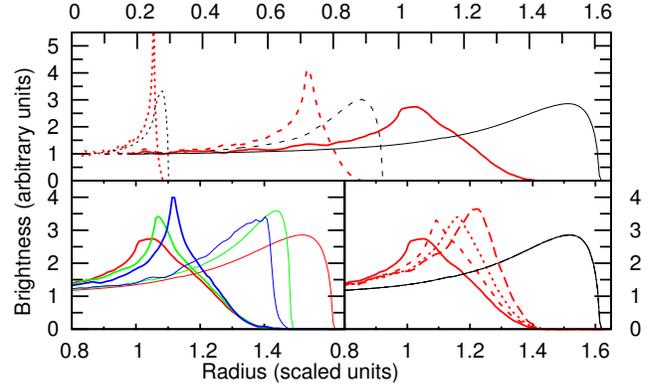}

\caption{\label{fig:bright-vs-rad}Emission intensity as a function of radius for selected
line-of-sight projections. All profiles have been scaled to the average brightness of the central
region ($R<0.2R_{\mathrm{FS}}$) of that image. \emph{Top}: intensity profiles for the $\gamma=5/3$
run. Peaking from left to right, $t=0.12$ (dotted line), $t=0.75$ (dashed line), and $t=2.0$
(solid line). The thin black lines trace out intensity due to shocked ISM.
\emph{Bottom left}: intensity profiles for the three runs at $t=2.0$. Peaking from left to right,
$\gamma=5/3$ (red line), $\gamma=4/3$ (green line), $\gamma=6/5$ (blue line). The thin lines trace
out emissivity of shocked ISM. \emph{Bottom right}: intensity profiles for the $\gamma=5/3$ run at
$t=2.0$, with ionization age cutoffs in place. Peaking from left to right, $\tau_{\mathrm{min}}=0.0$
(thick solid line), $\tau_{\mathrm{min}}=0.8$ (dotted line), $\tau_{\mathrm{min}}=1.6$ (thin solid
line), $\tau_{\mathrm{min}}=2.4$ (dashed line). The thin black line is intensity of shocked ISM.}
\end{figure}

The angle-averaged brightness profiles for selected data sets are shown in
figure~\ref{fig:bright-vs-rad}, illustrating the effects of time, compressibility, and an ionization
age cutoff on emission. The effects of the ejecta gradient mentioned previously are visible as the
plateau in the late-time curve in the top panel.

\begin{table*}
\caption{\label{tab:2d-interfaces}Interface radii and ratios, from 2-D projections and 3-D data}
\begin{tabular}{cccccccccccccc}
\hline
$\gamma$ & $t$ & $\hat{R}_{\mathrm{RS}}$ & $\hat{R}_{\mathrm{CD}}$ & $\hat{R}_{\mathrm{FS}}$ & $\hat{R}_{\mathrm{RS}}:\hat{R}_{\mathrm{FS}}$ & $\hat{R}_{\mathrm{CD}}:\hat{R}_{\mathrm{FS}}$ & $R_{\mathrm{RS}}$ & $R_{\mathrm{CD}}$ & $R_{\mathrm{FS}}$ & $R_{\mathrm{RS}}:R_{\mathrm{FS}}$ & $R_{\mathrm{CD}}:R_{\mathrm{FS}}$ & $\xi_{\mathrm{RS}}$ & $\xi_{\mathrm{CD}}$ \\
\hline
5/3 & 0.12 & 0.250 & 0.279 & 0.296 & 0.845 & 0.943 & 0.249 & 0.265 & 0.297 & 0.838 & 0.892 & <1\% & 6\% \\
\textquotedbl{} & 0.75 & 0.725 & 0.873 & 0.951 & 0.762 & 0.918 & 0.716 & 0.800 & 0.952 & 0.752 & 0.840 & 1\% & 9\% \\
\textquotedbl{} & 2.0 & 1.009 & 1.391 & 1.616 & 0.624 & 0.861 & 1.005 & 1.229 & 1.615 & 0.622 & 0.761 & <1\% & 13\% \\
4/3 & 0.19 & 0.334 & 0.371 & 0.377 & 0.886 & 0.984 & 0.333 & 0.351 & 0.377 & 0.883 & 0.931 & <1\% & 6\% \\
\textquotedbl{} & 0.75 & 0.737 & 0.863 & 0.888 & 0.830 & 0.972 & 0.734 & 0.795 & 0.887 & 0.828 & 0.896 & <1\% & 8\% \\
\textquotedbl{} & 2.0 & 1.074 & 1.378 & 1.484 & 0.724 & 0.929 & 1.060 & 1.228 & 1.484 & 0.714 & 0.827 & 1\% & 12\% \\
6/5 & 0.16 & 0.297 & 0.332 & 0.332 & 0.895 & 1.000 & 0.297 & 0.313 & 0.331 & 0.897 & 0.946 & <1\% & 6\% \\
\textquotedbl{} & 0.75 & 0.733 & 0.877 & 0.883 & 0.830 & 0.993 & 0.737 & 0.795 & 0.872 & 0.845 & 0.912 & -2\% & 9\% \\
\textquotedbl{} & 2.0 & 1.115 & 1.393 & 1.449 & 0.769 & 0.961 & 1.112 & 1.224 & 1.438 & 0.773 & 0.851 & -1\% & 13\% \\
\hline
\end{tabular}
\end{table*}

The radii for the three fluid discontinuities, averaged around the rim of each projection, are
gathered in table~\ref{tab:2d-interfaces}, based off of the preceding analysis of the line-of-sight
projections. In this table and for the rest of the paper, we use the notation of \citet{CC08}, in
which $\hat{R}_{\mathrm{CD}}$ denotes the projected radius of the contact discontinuity and
$R_{\mathrm{CD}}$ the average location of the discontinuity in three dimensions. Comparing
$R_{\mathrm{FS}}$ between the runs with $\gamma=5/3$ and $\gamma=4/3$, it is clear that increasing
the compressibility of the fluid has a significant effect on the overall size of the remnant at
similar dynamical times, just as with the one-dimensional runs presented in
figure~\ref{fig:1d-3d-compare-1}. The forward shock decelerates less than the reverse shock and the
contact discontinuity as the remnant evolves, and all ratios in Table~\ref{tab:2d-interfaces} drop
with time. Lowering the adiabatic index has the opposite effect, bringing the interfaces closer
together and raising the ratios in table~\ref{tab:2d-interfaces}.

From shock positions alone it may be difficult to distinguish between a younger remnant with less
efficient cosmic ray acceleration and a more evolved remnant with more efficient particle
acceleration. At roughly the age of Tycho's remnant, the ratio
$\hat{R}_{\mathrm{CD}}/\hat{R}_{\mathrm{FS}}$ for the $\gamma=5/3$ run is lower than observed in
Tycho, as is $\hat{R}_{\mathrm{RS}}/\hat{R}_{\mathrm{FS}}$. With the $\gamma=4/3$ simulation,
$\hat{R}_{\mathrm{CD}}/\hat{R}_{\mathrm{FS}}$ is close to that observed in Tycho, but the reverse
shock is too close to the forward shock. If the forward shock were more efficient at cosmic ray
acceleration than the reverse shock, the effective adiabatic index of shocked ejecta and shocked ISM
would differ, leading to varying compressibilities ahead of and behind the contact discontinuity.
Previous work offers support for such a situation. \citet{DEB00} provided evidence that low magnetic
fields lead to negligible acceleration of cosmic rays. In the absence of magnetic field
amplification by the reverse shock, the frozen-in magnetic field of the ejecta should be attentuated
by expansion, and therefore $\gamma_{\mathrm{eff}}$ should be approximately $5/3$. \citet{EDB05}
elaborated on the bijective relationship between magnetic field amplification and DSA, and while
studies have been done of magnetic field amplification at Tycho's forward shock \citep{VBK05}, we
are unaware of observations suggesting magnetic field amplification at Tycho's reverse shock;
\citet{HRD00} did, however, find evidence of efficient cosmic ray acceleration at the reverse shock
of 1E 0102.2-7219. Our conclusion echoes that of \citet{Warr05}, which found little evidence for
particle acceleration at the reverse shock of Tycho by comparing 1-D hydrodynamic models to the
discontinuity radii deterimined in that paper.

\subsection{Implications for Tycho and SN 1006}\label{sub:Dynamical-age}

The evolution of the exponential model is governed entirely by the scaled age $t$, so attempts to
compare the model to observations must determine the dynamical age of the remnant in question. In
the previous sections we have examined the effect of scaled time and adiabatic index on the
appearance and structure of our simulated remnants, including interface locations, power spectra,
and deceleration parameters. Now we will combine these results with observed quantities to estimate
the dynamical ages of Tycho and SN 1006, and to discuss the efficiency of cosmic ray acceleration in
both remnants. We consider the fleecy ejecta structures in the SNRs, dynamical quantities of the
ejecta, and the average radial locations of the fluid discontinuities in determining the age of the
two SNRs. We also use deceleration parameters in order to constrain the effective adiabatic index at
the forward shock.

Tycho's SNR was first observed in 1572, giving it (at time of writing) a real age of 439 years.
This corresponds to a scaled age of
$t=439$~yr/$T^{\prime}\approx1.77\,(M_{e}/\mch)^{-5/6}E_{51}^{1/2}n_{0}^{1/3}$. SN 1006, with a real
age of 1005 years, has a scaled age of $t=4.05\,(M_{e}/\mch)^{-5/6}E_{51}^{1/2}n_{0}^{1/3}$. Despite
the difference in age, our simulations strongly suggest that Tycho be dynamically older than SN
1006. The fleecy structures present in Tycho are, on average, larger in angular size than those
observed in SN 1006. In our simulated remnants and their projections, the angular size of the fleecy
ejecta structures is almost completely determined by the dynamical age of the remnant (see
figures~\ref{fig:3d-times}, \ref{fig:temporal-53-2d}, and \ref{fig:sn1006-times}); the size of the
structures depends only weakly, if at all, on the compressibility of the fluid
(figures~\ref{fig:3d-gammas} and \ref{fig:gamma-t-late-2d}).

One way to approach the dynamical age of the remnant is consideration of the motion of the
ejecta. Three quantities are relevant here: (i) the free expansion velocity of unshocked ejecta just
ahead of the reverse shock, (ii) the bulk radial motion of shocked ejecta near the reverse shock,
and (iii) the bulk motion of the ejecta at the contact discontinuity. 
\begin{enumerate}
  \item The expansion velocity at the reverse shock of SN 1006 is $7026\pm13$~\kms
($v=0.83\,(M_{e}/\mch)^{1/2}E_{51}^{-1/2}$) based on Doppler-shifted absorption lines through the
remnant \citep{HFB07} (the shape of the lines points to asymmetry in the remnant, however, so it is
uncertain if this velocity is representative of the reverse shock everywhere). The shock is unlikely
to be accelerating cosmic rays efficiently, for reasons outlined in the previous subsection; the
effective adiabatic index should therefore be close to $\gamma=5/3$. In this case, the location of
the reverse shock is consistent in one dimension and in three. From 1-D runs, then, we find that a
free expansion velocity of 0.83 at the reverse shock corresponds to a dynamical age of $t=1.0$ for
SN 1006 for canonical values of $E_{51}=1$ and $M_{e}=\mch$.
  \item No observations like those of \citet{HFB07} have been reported for Tycho, but
\citet{Hayato10} fitted pairs of red- and blue-shifted absorption lines to the Fe K$\alpha$ spectral
feature of Tycho, and measured the radial velocity by quantifying the relative shift between the
center of the remnant and the edge. They report expansion velocities of $4000\pm300$~\kms
($v=0.47\,(M_{e}/\mch)^{1/2}E_{51}^{-1/2}$) for Fe K$\alpha$, which should be associated with
freshly shocked ejecta at the reverse shock. The velocity of recently shocked ejecta in our
$\gamma=5/3$ simulation at $t=0.75$ ($t=2.0$) is $v=0.58$ ($v=0.21$). Interpolation between the two
suggests a dynamical age of 1.1 for Tycho.
  \item Using a similar process for the Si He$\beta$ feature, \citet{Hayato10} found the expansion
velocity to be $4700\pm100$~\kms ($v=0.56\,(M_{e}/\mch)^{1/2}E_{51}^{-1/2}$) for silicon, which they
associate with the contact discontinuity. In 1-D the radial velocity of the contact discontinuity is
0.56 at $t=0.72$. However, in multiple dimensions fluid instabilities cause a range of velocities
and densities in the mixing region. The greatest emissivities should occur in the forward tips of
Rayleigh-Taylor structures, which expand more rapidly than the rest of the contact discontinuity.
The highest ionization ages should also occur in the R-T structures. In the $\gamma=5/3$ run, the 10
per cent of the ejecta with the highest ionization age has a radial velocity of 0.65 at $t=0.75$;
this decreases to $v=0.26$ by $t=2.0$. Interpolating between the two numbers, the results of
\citet{Hayato10} applied at the contact discontinuity of our 3-D simulations imply that Tycho's
dynamical age is 1.0. 
\end{enumerate}
The estimated age around 1.0 for both Tycho and SN 1006 conflicts with the size of the ejecta
structures, which should be a marker for relative age. The reason for this difference is not clear.

We now assess the ramifications of observed fluid velocities on the parameters ($n_{0}$, $M_{e}$,
and $E_{51}$) of the two SNRs. As mentioned in section~\ref{sec:The-exponential-model}, the
scaling factor for time depends inversely on both the explosion energy and the interstellar density
(it also depends on the ejecta mass, but we see little reason why this should deviate from $\mch$).
There is a good deal of evidence that the ISM around Tycho is denser than that around SN 1006: the
inferred densities around Tycho from X-ray \citep{CC07}, $\gamma$-ray \citep{VBK08}, and optical
observations \citep{KWC87} all support a value for $n_{0}$ of 0.1-0.3~cm$^{-3}$ in the west, with
higher densities in the east. Assuming $E_{51}=1$, the scaled and physical ages of Tycho match if
$n_{0}=0.18$~cm$^{-3}$, consistent with observations.

\begin{table}
\caption{\label{tab:SNR-params}SNR 1006 parameters, assuming $M_{e}/\mch=1$}
\begin{tabular}{cccccc}
\hline
$M_{e}/\mch$ & $E_{51}$ & $v_{\mathrm{RS}}^{(1)}$ & $t^{(2)}$ & $n_{0}~(\mathrm{cm}^{-3})^{(3)}$ & $r^{\prime}_{\mathrm{FS}}~(\mathrm{pc})^{(4)}$ \\
\hline
1.0 & 1.0 & 0.83 & 0.98 & 0.014 & 8.84 \\
1.0 & 1.5 & 0.68 & 1.35 & 0.020 & 9.31 \\
1.0 & 3.0 & 0.48 & 2.15 & 0.029 & 10.42 \\
1.0 & 4.0 & 0.42 & 2.50 & 0.030 & 11.03 \\
1.0 & 6.0 & 0.34 & 3.18 & 0.031 & 12.04 \\
1.0 & 9.0 & 0.28 & 3.67 & 0.028 & 13.37 \\

1.5 & 2.0 & 0.72 & 1.25 & 0.029 & 9.13 \\
1.5 & 2.5 & 0.64 & 1.48 & 0.035 & 9.39 \\
1.5 & 3.0 & 0.59 & 1.66 & 0.037 & 9.69 \\
1.5 & 4.0 & 0.51 & 2.00 & 0.042 & 10.17 \\
1.5 & 6.0 & 0.42 & 2.50 & 0.043 & 11.03 \\
1.5 & 9.0 & 0.34 & 3.10 & 0.047 & 11.99 \\
\hline
\end{tabular}
\\(1) 7026 km~s$^{-1}$ in scaled units.
\\(2) Time at which reverse shock velocity equals $v_{RS}$.
\\(3) Density required for the scaled time $t$ to correspond to 1001 yr.
\\(4) At scaled time $t$.
\end{table}

Around SN 1006, observations suggest a lower interstellar density. \citet{ABD07} found the X-ray
emission measure of the forward shock and calculated an ISM density of 0.05~cm$^{-3}$ in the
southeast quadrant, and argued that the density was roughly constant everywhere except the
filamentary northwest rim. Using expansion data, \citet{Katsuda09} determined the density at the
northeast to be 0.085~cm$^{-3}$. A very low ISM density, $n_{0}=0.014$~cm$^{-3}$ if $E_{51}=1$, is
required to match the dynamical and physical ages (see table~\ref{tab:SNR-params}). This is lower
than either observational estimate, though it is within model-dependent uncertainties of the value
presented by \citet{ABD07}. Also visible in table~\ref{tab:SNR-params} is that the physical age of
SN 1006 cannot be matched to the velocity measurements of \citet{HFB07} for any ambient density
greater than about 0.030~cm$^{-3}$, which itself requires $t>2.15$ and $E_{51}>3.0$. We can impose
an additional constraint on the parameters of SN 1006 by considering its size: at a distance of
$2.18\pm0.08$~kpc \citep{Wink03}, and with an average radius of 14.5$^{\prime}$ in the SE quadrant
\citep{CC08}, the radius of the forward shock is 9.19~parsecs. Under the assumption that
$M_{e}/\mch=1$, the explosion energy $E_{51}=1.4$ and ISM density $n_{0}=0.019$ are consistent with
SN 1006's age, reverse shock velocity, and size. These parameters suggest a scaled age of 1.3 for SN
1006. If we relax the assumption that $M_{e}=\mch$, we again find a consistent solution.
Table~\ref{tab:SNR-params} shows the same calculations if $M_{e}/\mch=1.5$. It can be seen that the
upper limit on $n_{0}$ is 0.047~cm$^{-3}$ in this case. The parameters that match the age, size, and
reverse shock velocity are $E_{51}=2.1$ and $n_{0}=0.030$~cm$^{-3}$; they also imply that SN 1006's
dynamical age is 1.3. Such a massive and energetic supernova could be consistent with a white dwarf
merger origin for SN 1006.

The deceleration parameter, like the separation between fluid discontinuities and the free expansion
velocity of ejecta, is a monotonic function of time. As evidenced by
figure~\ref{fig:1d-3d-decel-param}, it is also sensitive to the effective adiabatic index: for any
particular time, decreasing the adiabatic index also lowers the deceleration parameter. Tycho's
deceleration parameter is higher along the NW-SW rim (where the remnant is close to spherical
symmetry), $0.59\pm0.12$ \citep{Katsuda10}, than is SN 1006's at the NE rim (the only quadrant where
such a study has been performed in X-rays), $0.54\pm0.06$ \citep{Katsuda09}, although the two
numbers are separated by a single standard deviation. Since the deceleration parameter should decay
from 1.0 to 0.4 as a remnant transitions from free expansion to the Sedov phase, these numbers
suggest that Tycho is slightly dynamically younger than SN 1006. The large uncertainties in both
values do leave open the possibility that the situation is reversed -- that the correct deceleration
parameter for Tycho is lower than that for SN 1006 -- but the data presented in
figure~\ref{fig:1d-3d-decel-param} offer another interpretation consistent with observations. We
posit that, if SN 1006 is dynamically younger than Tycho -- or even the same dynamical age -- it
must be more efficiently accelerating cosmic rays at its forward shock, lowering both its
deceleration parameter and effective adiabatic index below Tycho's.

To quantify the difference in adiabatic index of the two remnants, we turn to the forward shock and
its separation from the contact discontinuity. There is substantial evidence that both remnants are
efficiently accelerating cosmic rays at their forward shock, so we exclude the $\gamma=5/3$ run and
focus on the $\gamma=4/3$ and the $\gamma=6/5$ runs. At $t=1.0$,
$\hat{R}_{\mathrm{CD}}\,:\,\hat{R}_{\mathrm{FS}}$ is 0.96 and 0.99, respectively, for $\gamma=4/3$
and $\gamma=6/5$. This ratio for SN 1006 is 0.98 in the NE polar cap \citep{CC08}. The effective
adiabatic index in SN 1006 should therefore be just over 1.2 -- at the very least, much closer to
$\gamma=6/5$ than to $\gamma=4/3$. The estimated adiabatic index is less than $6/5$ if the age
estimate is increased to 1.3, as suggested by Table~\ref{tab:SNR-params}. In the case of Tycho,
$\hat{R}_{\mathrm{CD}}\,:\,\hat{R}_{\mathrm{FS}}$ along the western rim (where the remnant appears
mostly spherical) is around 0.95. We expect an effective adiabatic index of just over $4/3$ for
Tycho (this number increases if the dynamical age decreases, and vice versa). We note that these
numbers agree broadly with the results of \citet{KBV11}, who used 1-D simulations to calculate
a range of allowable compressibilities that could match the radial morphology of Tycho and SN 1006,
and found that the upper end of SN 1006's range was higher than the upper end of Tycho's.

As mentioned in section~\ref{sub:Dependence-on-time} and table~\ref{tab:2d-interfaces}, the reverse
shock and contact discontinuity are constantly receding relative to the forward shock; the separation
between the interfaces could therefore serve as a probe of dynamical age. No reverse shock has been
directly observed in SN 1006, but \citet{Warr05} identified Fe K$\alpha$ emission with the reverse
shock in Tycho. Since the location of the contact discontinuity is far less dependent on the
adiabatic index than are the locations of the forward and reverse shocks, 
$\hat{R}_{\mathrm{RS}}\,:\,\hat{R}_{\mathrm{CD}}$ requires fewer assumptions as an indicator of
Tycho's dynamical age than does the more commonly quoted figure of
$\hat{R}_{\mathrm{RS}}\,:\,\hat{R}_{\mathrm{FS}}$. For Tycho, this value is $0.73/0.96=0.76$
\citep{Warr05}. The same ratio for the $\gamma=5/3$ run at $t=0.75$ is 0.83, and has dropped to 0.73
by $t=2.0$. Interpolation between the two yields a dynamical age for Tycho of 1.6. The ratio
$\hat{R}_{\mathrm{RS}}\,:\,\hat{R}_{\mathrm{FS}}$ can be used to estimate Tycho's dynamical age, but
requires an assumption about the compressibility of the ejecta and ISM. Using $\gamma=5/3$ for the
ejecta (and $\hat{R}_{\mathrm{RS}}$) and $\gamma=4/3$ for the ISM (and $\hat{R}_{\mathrm{FS}}$), the
appropriate ratio is 0.82 at $t=0.75$ and 0.68 at $t=2.0$. The ratio in Tycho is 0.73
\citep{Warr05}, which occurs at $t=1.6$. This result does not change significantly if more
compressible ISM is assumed. Intriguingly, velocity measurements and radial morphology each give a
consistent estimate for the age of Tycho's remnant, but the two estimates do not agree with each
other.

Taken together, the results discussed in this section paint the following scenario for the relative
dynamical ages and effective adiabatic indices of the remnants of Tycho's SN and SN 1006. The size
of ejecta stuctures and Tycho's radial morphology suggests that it  is the older remnant, but
available velocity information implies that both remnants have roughly equal dynamical ages. In
either case, SN 1006 is more efficiently accelerating cosmic rays at its forward shock. We calculate
a dynamical age for SN 1006 of 1.3 based on the free expansion velocity of unshocked ejecta and on
the known size of the remnant. At this age, the separation between the forward shock and the contact
discontinuity implies an effective adiabatic index of $6/5$. If the scaled age of Tycho's SNR is 1.0
as suggested by expansion velocity of its ejecta, we find that $\gamma_{\mathrm{eff}}\approx4/3$ at
its forward shock. Given the simplicity of our model, the disparate age estimates in the case of
Tycho, and the numerous factors affecting the remnants' actual expansions, however, further study is
warranted before firm conclusions can be drawn.

\section{CONCLUSIONS}\label{sec:Conclusions}

We have performed high resolution three-dimensional simulations of a Type Ia supernova remnant using
an exponential ejecta profile and assuming homogenous ejecta and ISM. The Rayleigh-Taylor instability
is clearly capable of generating the fleecy structures observed in Tycho's SNR. Comparison against
SN 1006 is qualitatively less favorable due to the orientation and spacing of the structures in that
remnant. The ejecta and ISM in all simulations were smooth at initialization, implying that
clumpiness is not a necessary condition to generate the structures observed in Tycho. This result
depends critically on evolving the simulations long enough time for the instabilities to saturate,
independently of the compressibility of the fluid. While our simulations reproduced the central
regions of both remnants well, they require a relatively high cutoff in ionization age to capture
the indistinct limb brightening seen in Tycho, and qualitatively fail to match SN 1006.

After consideration of several observables tied to dynamical age -- such as expansion rate of ejecta
at the reverse shock, average radii of fluid discontinuities, and deceleration parameter for select
regions around the rim -- we find that observed parameters are inconsistent in the case of Tycho's
SNR, with radial structure and ejecta morphology presenting a different age ($t=1.6$) than observed ejecta
velocities ($t=1.0$). Taking the lower value as the more accurate, Tycho's effective adiabatic index
is slightly higher than $4/3$ at its forward shock. SN 1006's dynamical age is approximately 1.3,
and accurate measurements of its distance and size allow us to calculate $E_{51}=1.4$ and
$n_{0}=0.019$~cm$^{-3}$. Acceleration of cosmic rays at the forward shock of SN 1006 is more
efficient than at that of Tycho's SNR: $\gamma_{\mathrm{eff}}$ is $6/5$.

Although shape of the contact discontinuity depends on adiabatic index and compressibility of the
ejecta, its average radius does not for the period of time covered by our simulations. Further, the
shape of the CD, especially given the close proximity to the forward and reverse shocks at low
$\gamma$, affects the shape of the shock fronts. For the lowest value of $\gamma$ used, knots of
ejecta appear to protrude outside the forward shock, but in reality lie just inside faint bubbles of
emission from shocked ISM. This does not happen on angular scales as large as the shelf of emission
in NE Tycho or the polar region of emission in NE SN 1006. Since instabilities and smooth ejecta
cannot generate those features, something else (inhomogeneity in ISM, ambient magnetic fields, or
asymmetric explosion) must be necessary. Studies including inhomogenous ejecta or an ambient
magnetic field to azimuthally affect cosmic ray production offer additional insight into the
large-scale structure of both of these remnants.

\section*{Acknowledgements}
We thank S.P. Reynolds and K.J. Borkowski for valuable discussion, and the anonymous referee for
comments that improved the clarity of the paper. Work for this paper was performed under the GAANN
Fellowship and NSF grants TG-MCA08X010 and AST-0708224. We also acknowledge the Texas Advanced
Computing Center (TACC) at The University of Texas at Austin for providing HPC resources that have
contributed to the research results reported within this paper. URL: http://www.tacc.utexas.edu.

\appendix
\section{DIFFERENCES BETWEEN 2-D PROJECTIONS AND 3-D DATA}\label{app:proj-effects}

Projections of an irregular three-dimensional spherical structure are biased towards protrusions
from the average radius, so the average radius of the contact discontinuity differs from the
observed radius in projection. This bias can be parameterized by a correction factor
$\xi_{\mathrm{proj}}$, where $R_{\mathrm{proj}}=(1+\xi_{\mathrm{proj}})R_{\mathrm{true}}$. The
correction factor depends strongly on the degree of structure present in a surface, so the forward
and reverse shocks of our model remnants should see minimal correction compared to the contact
discontinuity. In \citet{Warr05} several factors (e.g. the observed power spectrum and expected
length of Rayleigh-Taylor fingers) were considered in determining a correction factor of 6\% for
Tycho, and resulted in a revision from observed ratios (1~:~0.96~:~0.73) to ``true'' ratios
(1~:~0.93~:~0.71). Drawing on the work of \citet{DC98} and \citet{WC01}, \citet{CC08} arrived at a
projection correcting factor of 10\% for SN 1006. With access to both projections and fully
three-dimensional data, we can directly evaluate the errors of this method.

The projection correcting factor is related to the projected and true discontinuity locations by
$(\hat{R}_{\mathrm{CD}}/\hat{R}_{\mathrm{FS}})=(R_{\mathrm{CD}}/R_{\mathrm{FS}})\cdot(1+\xi)$; a
similar equation exists for the reverse shock. Table~\ref{tab:2d-interfaces}, in addition to the
interface locations in the projections, also lists both the true discontinuity locations and the
projection correction factor for each $\gamma/t$ pair. Both the CD/FS and RS/FS ratios are higher in
projection than they are in 3-D for every simulation at almost every time; the two exceptions for
$\gamma=6/5$ are likely caused by the filamentary nature of emission at the forward shock and the
extensive interaction between the fluid discontinuities. Additionally, $\xi$ is very nearly 0 for
the reverse shocks. This result is in keeping with the relative smoothness of both the forward and
reverse shocks as seen in figures~\ref{fig:3d-times} and \ref{fig:3d-gammas}. \citet{Warr05}
predicted a $\xi_{\mathrm{proj}}$ of 3\% for the reverse shock of Tycho, an overestimation of the
structure present in the reverse shock.

For the contact discontinuity, the correction factors range from 6\% at the earliest times to
12-13\% at the latest, showing remarkable consistency across $\gamma$ despite the changes to the
shape of both the forward shock and the contact discontinuity. \citet{CC08} suggested that a $\xi$
upwards of 10\% would be necessary to match observations of SN 1006
($R_{\mathrm{CD}}/R_{\mathrm{FS}}=0.96$ for the SE quadrant where minimal nonthermal emission is
seen) with their one-dimensional simulations. Their runs were carried out to a scaled time of
$t\approx1.2$, depending inversely on the interstellar density in the environment of SN 1006. Given
the lack of synchrotron emission observed over this region, for our hydrodynamic models to match
that shock ratio we would require an exceptionally low ambient density in those regions (reducing
the dynamic age of the remnant) in conjunction with efficient acceleration.
\label{lastpage}

\end{document}